\documentclass[a4paper,fleqn,usenatbib]{mnras}

\usepackage{newtxtext,newtxmath}

\usepackage[T1]{fontenc}

\usepackage{graphicx}	
\usepackage{subfig}

\newcommand{\ds}{$\,{\rm deg}^2$} 

\title[SF to AGN transition at $\mathit{z \simeq 4}$]{The rapid transition from star-formation to AGN dominated rest-frame UV light at $\mathbf{z \simeq 4}$} 

\author[R. A. A. Bowler et al.]{R. A. A. Bowler,$^{1}$\thanks{E-mail: rebecca.bowler@physics.ox.ac.uk}
N. J. Adams$^{1}$,
M. J. Jarvis$^{1, 2}$
B.~H{\"a}u\ss ler$^{3}$
\\
$^{1}$Department of Astrophysics, University of Oxford, The Denys Wilkinson Building, Keble Road, Oxford, OX1 3RH \\
$^{2}$Department of Physics, University of the Western Cape, Bellville 7535, South Africa\\
$^{3}$European Southern Observatory, Alonso de Cordova 3107, Vitacura, Santiago, Chile }

\pubyear{2020}

\begin{document}

\label{firstpage}
\pagerange{\pageref{firstpage}--\pageref{lastpage}}
\maketitle

\begin{abstract}
With the advent of deep optical-to-near-infrared extragalactic imaging on the degree scale, samples of high-redshift sources are being selected that contain both bright star-forming (SF) galaxies and faint active galactic nuclei (AGN).
In this study we investigate the transition between SF and AGN-dominated systems at $z \simeq 4$ in the rest-frame UV. 
We find a rapid transition to AGN-dominated sources bright-ward of $M_{\rm UV} \simeq -23.2$.
The effect is observed in the rest-frame UV morphology and size-luminosity relation, where extended clumpy systems become point-source dominated, and also in the available spectra for the sample.
These results allow us to derive the rest-frame UV luminosity function for the SF and AGN-dominated sub-samples.
We find the SF-dominated LF is best fit with a double-power law, with a lensed Schechter function being unable to explain the existence of extremely luminous SF galaxies at $M_{\rm UV} \simeq -23.5$. 
If we identify AGN-dominated sources according to a point-source morphology criterion we recover the relatively flat faint-end slope of the AGN LF determined in previous studies.
If we instead separate the LF according to the current spectroscopic AGN fraction, we find a steeper faint-end slope of $\alpha = -1.83 \pm 0.11$.
Using a simple model to predict the rest-frame AGN LF from the $z = 4 $ galaxy LF we find that the increasing impact of host galaxy light on the measured morphology of faint AGN can explain our observations.

\end{abstract}




\section{Introduction}
How supermassive black holes and their host galaxies co-evolve over cosmic time poses many fundamental questions within Astrophysics.
The detection of luminous quasars at very high redshift (e.g.~\citealp{Fan2003, Willott2010, Mortlock2011, Banados2016, Banados2018, Yang2020}) demonstrates that active black holes are present less than a Gyr after the Big Bang.
Within the same epoch, the star-forming (SF) galaxy population is known to be building-up rapidly from measurements of the evolving rest-frame UV luminosity function (LF; e.g.~\citealp{Bouwens2015, Finkelstein2015, Bowler2020, Ono2018}).
Until recently, the populations of quasars and SF galaxies at redshifts $z = 4-8$ have typically been treated as separate, due primarily to the disparate luminosity space occupied by the current samples.
This is despite the majority of galaxies and quasars at very high-redshifts being selected based on the same spectral feature in optical/NIR survey data; the Lyman-continuum and/or the Lyman-$\alpha$ break.
The strong Lyman-break in the spectral energy distribution (SED), which is redshifted into the optical filters at $z \gtrsim 3$ and near-IR at $z \gtrsim 7$, has allowed large samples of UV-bright galaxies and AGN\footnote{In this work we use the more inclusive term AGN rather than quasar throughout.} to be selected efficiently.
In the last decade, the advent of intermediate surveys that probe areas up to a few hundred square degrees on the sky has led to the first samples that bridge both faint AGN as well as bright galaxies, filling in a previously unachievable parameter space in volume and luminosity~\citep{Matute2013, Kashikawa2014a, Matsuoka2018a, Stevans2018, Ono2018, Adams2020}.
The properties of these intermediate luminosity sources are important for several reasons.
Firstly, the existence of very UV bright, highly star-forming galaxies can challenge models of feedback and dust obscuration via the inferred steepness of the bright-end of the Lyman-break galaxy (LBG) UV LF (e.g.~\citealp{Bower2012, GonzalezPerez2013, Bowler2014, Dayal2014, Clay2015}).
Furthermore, the uncertainty in the number of the brightest galaxies in combination with `contamination' of these samples with faint-AGN or interloper populations can confuse the interpretation of the shape and evolution of the galaxy LF (e.g.~\citealp{Bowler2012, Bian2013}).
Secondly, the determination of the faint-end slope of the AGN LF is crucial for understanding if these sources played any significant role in reionizing the Universe at $z \gtrsim 7$ (e.g. as advocated by~\citealp{Giallongo2015, Giallongo2019}, see discussion in~\citealp{Parsa2018}).
Thirdly, samples of sources in which the AGN and stellar component both contribute measurably to the observed light give an insight into how and when black holes become intricately linked to their host galaxy (e.g. via measurements of the black-hole to bulge/stellar mass relation at very high-redshift;~\citealp{Willott2010a, Venemans2017}).

While the rest-frame UV $z \simeq 4$ AGN LF was first measured several decades ago (e.g.~\citealp{Warren1994, Richards2006, Masters2012, Ikeda2012}), recent surveys have been able to select larger samples over a wider luminosity range, thus providing greater precision.
In particular, there have been several successful campaigns to identify fainter sources at $z \ge 4$ with surveys such as the Subaru High-$z$ Exploration of Low-Luminosity Quasars (SHELLQs:~\citealp{Matsuoka2018}), the Infrared Medium Survey (IMS:~\citealp{Kim2019}) and the Hyper-SuprimeCam Stragetic Survey Program (HSC-SSP;~\citealp{Akiyama2018}).
These studies have been able to constrain the faint-end\footnote{The AGN LF is typically parameterised as a double-power law (DPL) of the form $\phi \propto \phi^{*}\,/((L/L^*)^{\alpha} + (L/L^*)^{\beta})$.  This functional form includes four free parameters, a bright and faint-end slope ($\beta$ and $\alpha$), a characteristic luminosity ($L^*$) and normalisation $\phi^*$.  For fitting the rest-frame UV LF of LBGs a Schechter function is commonly assumed of the form $\phi \propto \phi^{*}\,(L/L^*)^{\alpha}\,e^{-L/L^*}$.  For the Schechter function the bright-end slope is replaced by an exponential decline bright-ward of $L^*$.} of the AGN LF for the first time at $z \gtrsim 4$, however there remain large discrepancies in the derived faint-end slope which ranges from $\alpha \simeq -1.3$~\citep{Matsuoka2018a, Akiyama2018} to as steep as $\alpha \simeq -2$~\citep{McGreer2018, Giallongo2019, Shin2020}.
A key challenge in the robust determination of the number density of the lowest luminosity AGN is that faint-ward of a certain absolute UV magnitude ($M_{\rm UV} \simeq -23$;~\citealp{Stevans2018, Ono2018, Adams2020}), LBGs become overwhelmingly more numerous.
In response, AGN selection methodologies have typically included a condition that the source must be unresolved in imaging data, as expected for a source dominated by the central AGN.
Despite this, the very faintest sources targeted by these studies have shown spectra that are typical of Lyman-break galaxies (e.g.~\citealp{Matsuoka2018a, Kashikawa2014a}).
Furthermore, as studies probe fainter AGN in the rest-frame UV, it is unclear how the light from star-formation might impact the morphology (e.g.~\citealp{Gavignaud2006}) and hence cause current selection procedures based on compactness to become incomplete.
Thus a more detailed analysis of the properties of faint-AGN and bright galaxies is required.

In~\citet{Adams2020} we selected a sample of LBGs and AGN at $z \simeq 4$ from the COSMOS and XMM-LSS deep extragalactic fields using a photometric redshift analysis based on the ground-based optical to NIR photometry.
The advantages of this sample over previous studies are i) we do not impose any condition on source size or morphology and hence we are complete to both point-sources and extended galaxies, ii) we exploit the NIR data which results in a very clean selection of $z \simeq 4 $ sources and iii) we have used two of the most widely studied deep fields where there is wealth of deep multi-wavelength data and spectroscopy available.
Here we utilise this sample to investigate the properties of objects within the `transition' regime between bright-AGN and the typical galaxy population at this redshift.
We do this by looking at the morphology and size of the sources using both the available ground-based and wide-area~\emph{Hubble Space Telescope} (\emph{HST}) mosaics in COSMOS.
In addition we have compiled publicly available spectra for the sample and use this to further classify sources.
The structure of the paper is as follows.
In Section~\ref{sect:data} we describe the variety of datasets we use, and in Section~\ref{sect:size} we describe the size measurements and the results from the available archival spectra of the sample.
In Section~\ref{sect:lf} we derive the AGN fraction from our data and estimate the separated SF and AGN-dominated LFs.
We discuss our results in~\ref{sect:dis} and present a simple empirical model of the AGN LF which we use to interpret our results in Section~\ref{sect:model}.
We end with conclusions in~\ref{sect:conc}.
Throughout this work we present magnitudes in the AB system~\citep{Oke1974,Oke1983}.
The standard concordance cosmology is assumed, with $H_{0} = 70 \, {\rm km}\,{\rm s}^{-1}\,{\rm Mpc}^{-1}$, $\Omega_{\rm m} = 0.3$ and $\Omega_{\Lambda} = 0.7$.
At $z = [3.5, 4.0, 4.5] $ this cosmology implies that one {\rm arcsec} corresponds to physical distances of $[7.3, 7.0, 6.6]\,{\rm kpc}$.

\section{Sample selection and data}\label{sect:data}

The sample of $z \simeq 4$ galaxies and AGN we utilize in this work was selected in the COSMOS and XMM-LSS deep extragalactic fields.
The selection was based on a photometric redshift fitting of the optical to NIR bands (u-band to $K_s$) from the available ground-based data.
In this paper we further include~\emph{HST} ACS imaging and spectroscopic observations available in the public domain. 

\subsection{The sample}
The full sample of $z \simeq 4$ sources from~\citet{Adams2020} consisted of 20064 (38722) sources in the COSMOS (XMM-LSS) fields bright-ward of the 50 percent completeness limit of $M_{\rm UV} \simeq -20$.
To be included in the sample the object must have a best-fitting photometric redshift in the range $3.5 < z < 4.5$ with either a galaxy or AGN SED.
Stars were removed using a relative $\chi^2$ cut, such that the galaxy or AGN template must have a better fit than the stellar model.
In this work we refined this sample using the most up-to-date photometry in the fields.
We matched the original catalogue with a new $I$-band selected catalogue (created in an identical fashion) that included the deeper HSC DR2 data, and the UltraVISTA DR4 imaging.
The matching process revealed a small number of artefacts that were removed in the newer HSC release.
We also required the objects to satisfy the selection criterion described above when run on the new deeper photometry, and be brighter than $15\sigma$ in the ground-based $I$-band to ensure a robust morphology analysis.
As a result of these steps we were left with a sample of 15126 (25592) sources in COSMOS (XMM-LSS). 
For the morphology analysis we required the sources to be covered by the~\emph{HST}/ACS mosaic in the COSMOS field.
The sub-sample covered by this mosaic contained 13848 sources.

\subsection{Imaging data}
Within the COSMOS field we used the~\emph{HST}/Advanced Camera for Surveys (ACS) mosaic in the F814W filter (hereafter $I_{814}$:~\citealp{Koekemoer2007,Scoville2007,Massey2010}).
We obtained $10 \times 10 \,{\rm arcsec}^2$ cut-outs\footnote{\url{https://irsa.ipac.caltech.edu/data/COSMOS/index_cutouts.html}} from this mosaic for each source in our COSMOS sample where the ACS data existed (corresponding to 96 percent of the sample).
The ACS data has a pixel scale of $0.03\,{\rm arcsec}/{\rm pix}$, and a typical point-source full-width-at-half-maximum (FWHM) of less than 0.1 arcsec.
The COSMOS field is uniformly covered to single orbit depth, leading to a $5\sigma$ depth of 27.0 (0.6 arcsecond diameter aperture).
Both the COSMOS and XMM-LSS fields contain a wealth of data in the optical and NIR bands.
In this study we measure the rest-frame UV size from the HSC $I$ and CFHT $i$-bands.
This allowed us to identify any potential systematics in the size measurement from using data of different depth and seeing.
The HSC $I$-band is available across both fields, with varying depth, while the CFHT $i$-band is uniform in depth but is available for only a 1\ds~subsection of each field~\citep{Bowler2020}.
The images have a pixel scale of $0.15$ and $0.2$ in COSMOS and XMM-LSS respectively.
The seeing in these bands was approximately 0.65 arcsec.

\subsection{Publicly available spectroscopy}\label{sect:spectroscopy}

\begin{figure}
\includegraphics[width =0.48\textwidth]{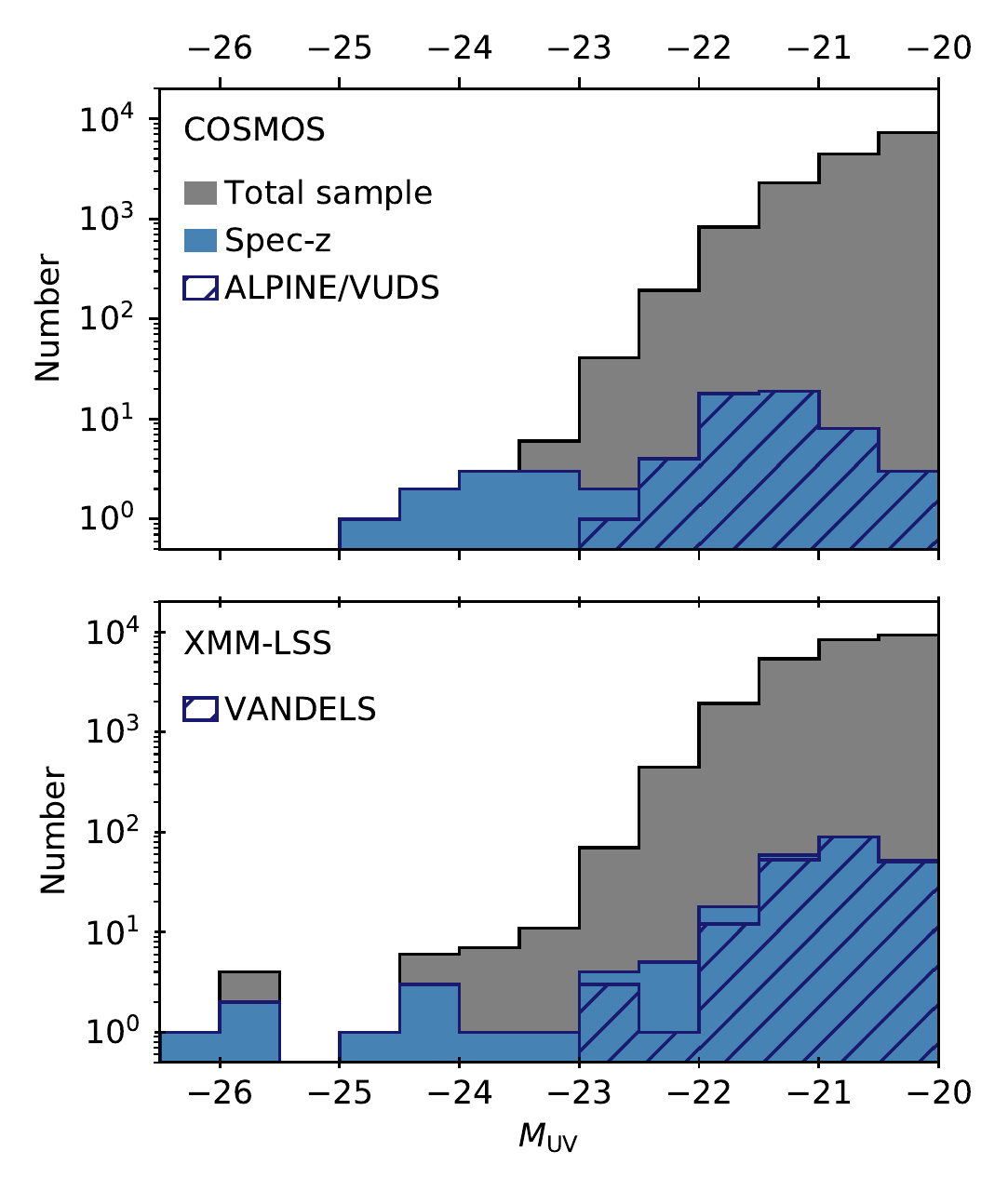}

\caption{The absolute UV magnitude distribution of our sample of $z \simeq 4$ sources selected initially in~\citet{Adams2020}.
The samples in the COSMOS and XMM-LSS fields are shown in the upper and lower plots respectively.
The total sample of galaxies and AGN with best-fit photometric redshifts in the range $3.5 < z < 4.5$ are shown as the grey histogram.
The spectroscopically confirmed sources are shown as the blue shaded histogram.
In hatched blue we have highlighted the spectroscopic redshifts from the deep galaxy surveys of ALPINE and VUDS in COSMOS and VANDELS in XMM-LSS.
The spectroscopic redshifts at brighter magnitudes are typically from magnitude limited surveys (e.g. SDSS, zCOSMOS; see Table~\ref{table:spec}).
}
\label{fig:hist}
\end{figure}

We endeavoured to extract the publicly available spectroscopy for the sample.
In both fields we initially matched to the compilation of spectroscopic redshifts created by the HSC team\footnote{\url{https://hsc-release.mtk.nao.ac.jp/doc/index.php/dr1_specz/}  Note that in this compilation, when a source had multiple redshifts from different surveys these were averaged.  This can result in a stated redshift at $z< 3$ when the source is securely at $z > 3$.  We corrected for this on a case-by-case basis.
In addition we removed redshifts from 3D-HST as these are not purely spectroscopic, particularly at higher redshifts where there are few spectral features in the rest-frame optical.
}, which we supplemented with additional catalogues from the VANDELS~\citep{Pentericci2018a,McLure2018} in XMM-LSS, and the ALMA Large Program to INvestigate (ALPINE;~\citealp{Fevre2019}) and the~\citet{Boutsia2018} sample of $z \simeq 4$ AGN in COSMOS.
In Fig.~\ref{fig:hist} we show the spectroscopically confirmed sources in comparison to the full sample as a function of absolute UV magnitude.
In total we found a total of 63 and 236 high-redshift sources with secure spectroscopic flags in COSMOS and XMM-LSS respectively (76 and 270 with all flags)\footnote{Secure flags were typically 3 or 4, including flags that may have been modified for the presence of AGN (e.g. a flag of 14 in VVDS denotes a secure AGN)}.
As part of this process we identified and removed 4 (12) low-redshift interlopers in COSMOS (XMM-LSS).
In COSMOS, 100 percent of the sources in our sample at $M_{\rm UV} < -23.5$ are confirmed spectroscopically, partially due to the campaign of~\citet{Boutsia2018}.
In XMM, we find a lower percentage of 42 percent in the same magnitude range.
At the bright-end of our sample the spectroscopic redshifts come primarily from magnitude limited surveys including the Sloan Digital Sky Survey (SDSS;~\citealp{Eisenstein2011,Richards2002}), zCOSMOS~\citep{Lilly2007}, VIMOS VLT Deep Survey (VVDS;~\citealp{LeFevre2015}) and Primus~\citep{Coil2011}.
At the faint-end of our survey the redshifts were obtained as specific follow-up for high-redshift galaxies.
The ALPINE sample includes sources from the VIMOS Ultra Deep Survey (VUDS;~\citealp{LeFevre2015}) and Deep Imaging Multi-Object Spectrograph (DEIMOS;~\citealp{Hasinger2018}) follow-up of high-redshift sources that are spread over the COSMOS field.
The VANDELS survey on the other hand, was limited to a smaller region of the field that overlaps with the~\emph{HST} Cosmic Assembly NIR Deep Extragalactic Legacy Survey (CANDELS;~\citealp{Grogin2011, Koekemoer2011}) where extremely deep spectroscopic integrations were performed.
Thus the VANDELS sources extend to fainter magnitudes than in ALPINE.
At the faint-end of the survey $\lesssim 2$ percent of our sample have been confirmed as part of these deep spectroscopic surveys.
While we cross-matched our sample to all available spectroscopic redshifts, we were only able to obtain a subset of reduced spectra depending on the survey.
We extracted all of the publicly available spectra from SDSS, zCOSMOS, VVDS and VANDELS for further analysis.

\section{Results}\label{sect:size}

Armed with our sample of $ z \simeq 4$ sources in the COSMOS and XMM-LSS fields, we proceeded to measure their sizes and spectroscopic properties.
As we are primarily concerned with the objects in the `transition' regime where AGN and LBGs have similar number densities, this analysis focuses on the results at $M_{\rm UV} \lesssim -22$.
At these bright magnitudes we have a larger proportion of spectroscopic follow-up from magnitude limited surveys and we are able to identify the source morphology at high S/N in the high-resolution~\emph{HST} data (e.g. even if the source fragments into several clumps the components are detected individually at $> 5\,\sigma$).

\begin{figure*}
\includegraphics[width =\textwidth, trim = 2.0cm 1.5cm 1.5cm 2cm]{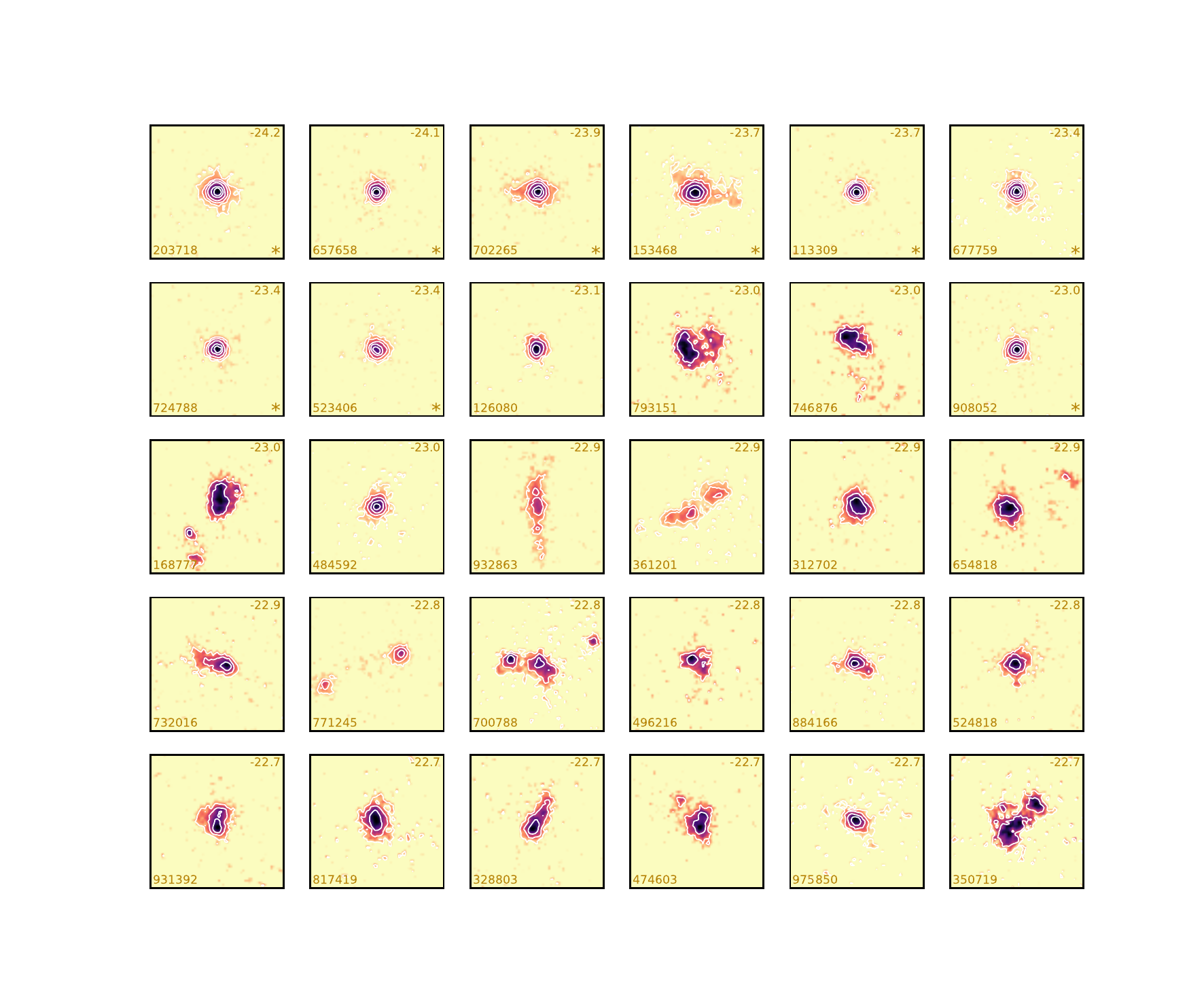}
\caption{Postage-stamp images in the~\emph{HSC}/ACS $I_{814}$ band of the brightest 30 sources in our COSMOS $z \simeq 4$ sample.
The sources are presented in order of $M_{\rm UV}$, with the brightest sources in the top left, spanning the range $-24.18 < M_{\rm UV} < -22.67$.
The stamps are $2\,{\rm arcsec}$ on a side (corresponding to $\sim 14\,{\rm kpc}$ at $ z = 4$), in the standard orientation of North to the top, and East to the left.
The images have been scaled by surface brightness, from $2\sigma$ (approximately $24 {\rm mag}/{\rm arcsec}$) to the peak.
Contours have been added at intervals of $1 {\rm mag}$ starting at the peak, to highlight the central compactness of the sources.
The ID number is shown in the bottom left and the $M_{\rm UV}$ is shown in the upper right.
Sources that have been spectroscopically confirmed are labelled with an astericks in the bottom right corner.
Note that all of the spectroscopically confirmed sources in this figure show strong quasar features in the spectra.
}
\label{fig:acs}
\end{figure*}

\subsection{Visual morphology}\label{sect:vis}

We first visually inspected the $z \simeq 4$ sources that had high-resolution imaging from the COSMOS~\emph{HST}/ACS $I_{814}$ mosaic.
In Fig.~\ref{fig:acs} we show postage-stamp images of the brightest 30 sources in our sample.
The brightest eight sources have been spectroscopically confirmed as quasars (see Table~\ref{table:spec}) and as expected, these sources appear compact.
As we go fainter in this sub-sample there is a dramatic change in the visual morphology, with the appearance of extended, clumpy, sources.
For example objects ID793151, ID746876 and ID168777 are clearly resolved, with multiple-components that extend $> 0.5\,{\rm arcsec}$ ($> 3.5 \,{\rm kpc}$) from the centroid.
This is as expected from the galaxy size-luminosity relation and extensive studies of similarly luminous sources at $z \simeq 3$ (e.g.~\citealp{Law2012, Lotz2006}) and $z > 5$ (e.g.~\citealp{Jiang2013b, Bowler2017}).
In the sources that are confirmed as AGN from their spectra, there is some evidence for weak extended emission (e.g. ID702265), which could be arising from the host galaxy.
Furthermore, one of the sources (ID153468) that is a confirmed quasar from the available rest-frame UV spectrum in~\citet{Boutsia2018}, appears to be compact but resolved (which is also confirmed by quantitative measure of the size, see Section~\ref{sect:agnfrac}).
This demonstrates that the host galaxy light is contributing significantly to the rest-frame UV light from this source.
We estimated the contribution from the host galaxy by simultaneously fitting a PSF and S{\'e}rsic profile (fixed to $n = 1.0$;~\citealp{Law2012}) to the image using {\sc GALFIT}.
The best-fit from this crude estimate was an equal contribution from the AGN and SF in this object.
For the other visual point-sources in the data, we found that a PSF alone was adequate to fit the imaging.
If we assume that the extended sources are SF-dominated, whilst the point-sources are AGN dominated (which is likely given the high rate of AGN spectra found for these sources), then we see a rapid transition in the individual galaxy morphology that indicates a transition from AGN to SF dominated systems at $M_{\rm UV} \simeq -23$.
A transition at this magnitude is in good agreement with that predicted by the simultaneous AGN and LBG LF fitting of~\citet{Adams2020} and we compare our results directly to this study in our analysis of the AGN fraction in Section~\ref{sect:agnfrac}.

\subsection{Size-luminosity relation}\label{sect:sl}

Due to the clumpy nature of the sources in the~\emph{HST} $I_{814}$ data, the sizes of these sources can be substantially biased depending on the chosen measurement technique.
When running {\sc Source Extractor} ({\sc SE};~\citealp{Bertin1996}) on these images for example, we found that the majority of the SF-dominated sources were de-blended into several component.
This results in a severe underestimate of the individual sizes of the brightest galaxies in the sample.
To get around this issue we proceeded to make high signal-to-noise stacks of the data over a range in $M_{\rm UV}$.

\subsubsection{Stacking procedure}
We created stacks in both the high-resolution~\emph{HST} imaging and the ground-based HSC and CFHT $i$-band data.
In both cases masks were formed from the {\sc SEGMENTATION} images created with {\sc SE}.
For the ground-based data we masked all sources that were not associated with the central source.
In the~\emph{HST}/ACS data we recombined de-blended components by retaining all objects that were within a radius of 0.8 arcsec from the central coordinate defined by the ground-based centroid.
Due to their close proximity, and the extended emission connecting clumps in many cases, we are confident these components are at the same redshift (see Fig.~\ref{fig:acs} and discussion in~\citealp{Bowler2017}).
If the separate components were galaxies at lower redshifts, we would expect these interloper sources to effect the ground-based optical to NIR photometry and thus be removed as interlopers by our SED fitting process.
With our recombined source we then determined the new centroid of the detected pixels in this extended object as the barycentre or first order moment (as used in {\sc SE}).
With these masked and centred images we proceeded to stack the images using both an average and median stack for comparison.
The size measurement we used was the half-light radius from {\sc SE}.
In the following plots we present the results derived from the median stack using the barycentre centroid, however we comment on any different results found using the other methods.
We obtained errors on our stacked size measurements using bootstrap resampling.
At the bright-end where there are very few sources in each bin this essentially measures the spread of the individual sizes in that bin.

\subsubsection{Results}
\begin{figure}
\begin{center}
\includegraphics[width =0.47\textwidth]{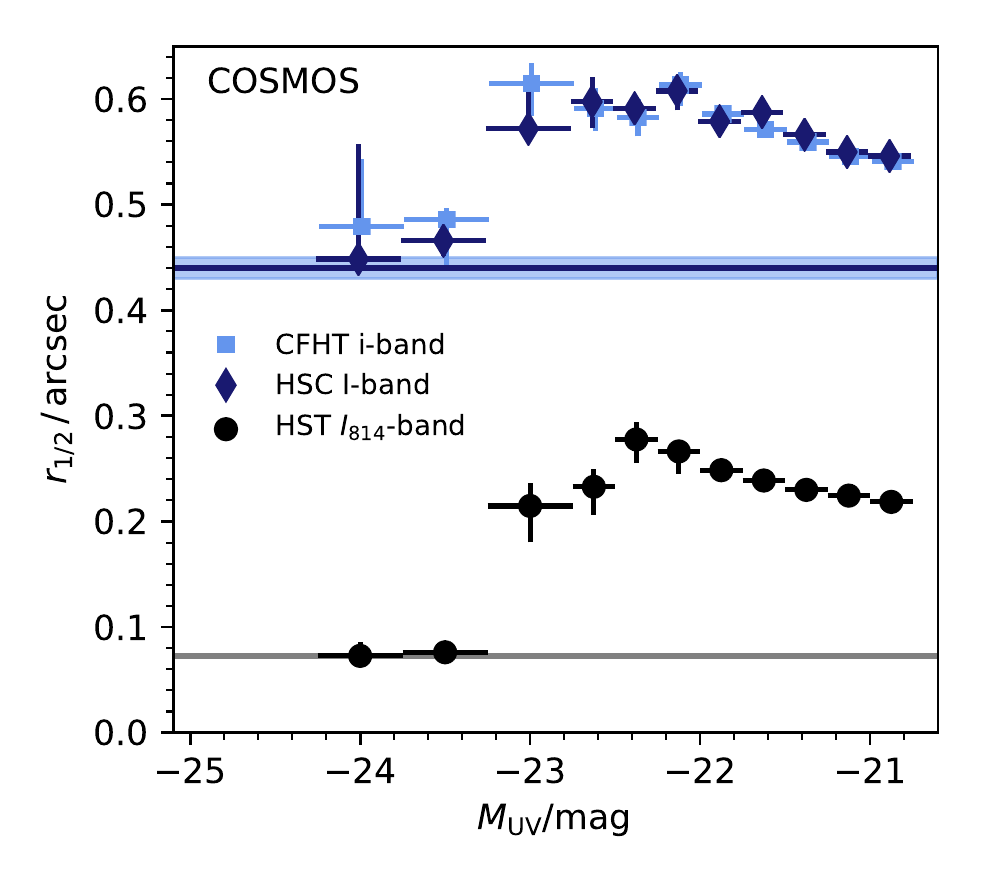}\\
\includegraphics[width =0.47\textwidth]{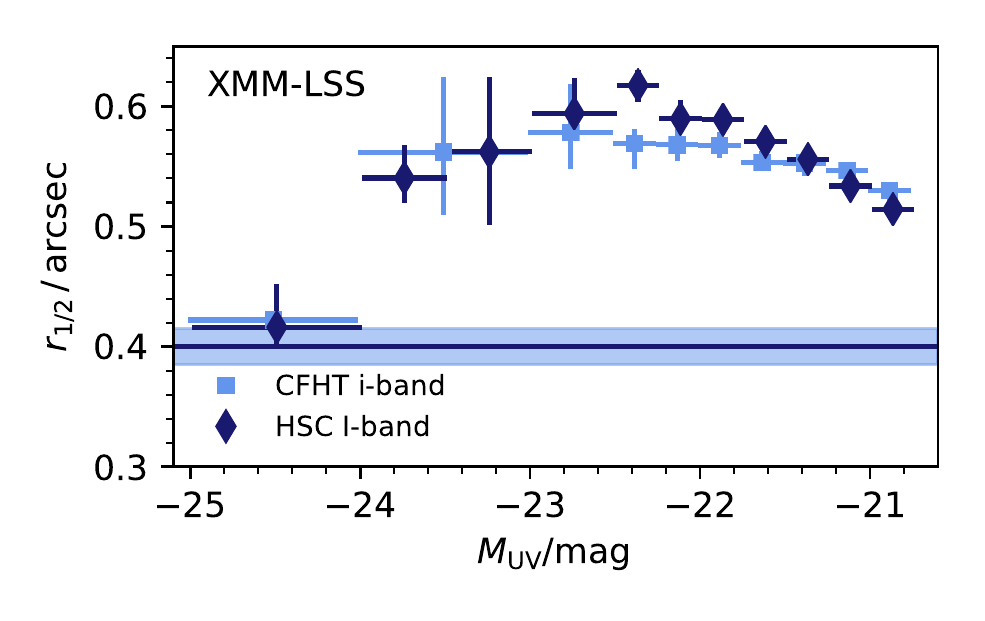}\\
\caption{The observed half-light radius of our sample of $z \simeq 4$ sources, uncorrected for the PSF, in the COSMOS and XMM-LSS fields (upper and lower plot respectively).
In each plot the dark blue diamonds show the HSC I-band measurement, and the light blue squares show the CFHT i-band result.
In the COSMOS plot we show the measured $r_{\rm 1/2}$ from the HSC $I_{814}$-band as the black circles.
The blue horizontal band shows the $r_{\rm 1/2}$ measured for the ground-based PSF.
The corresponding band for the ACS data is shown as the black/grey line.
In all datasets we see a drop off in measured size at $M_{\rm UV} < -22.5$, although this is less pronounced in XMM-LSS.}
\label{fig:sl}
\end{center}
\end{figure}

In Fig.~\ref{fig:sl} we show the observed sizes of the stack galaxy images, uncorrected for the PSF, as a function of $M_{\rm UV}$ for the ground-based and~\emph{HST} data.
At the faint-end of our sample we find that the galaxy stacks are resolved even in the ground-based data, with measured half-light radii in the range $r_{\rm 1/2} \simeq 0.55$--$0.6\,{\rm arcsec}$ as compared to $r_{\rm 1/2} \simeq 0.4$--$0.45\,{\rm arcsec}$ for the PSF.
As we move to brighter galaxy stacks there is a gentle increase in size until $M_{\rm UV} \simeq -22.5$, where we see a drop to smaller sizes that are consistent with being unresolved.
The results from the HSC $I$-band and the CFHT $i$-band are consistent within the errors for both fields.
In COSMOS, where we also have the higher resolution imaging~\emph{HST} $I_{814}$ data, the drop in size observed in the ground-based and~\emph{HST} data occurs at a consistent $M_{\rm UV}$.
The wider area covered by XMM-LSS can also explain the shallower decline in size bright-ward of $M_{\rm UV} = -23$ as compared to the COSMOS result, because this larger volume will result in rarer bright galaxies being detected.
Indeed, we identify a spectroscopically confirmed galaxy at $M_{\rm UV} = -23.6$ (see Section~\ref{sect:spectra}).
In both COSMOS and XMM-LSS, the brightest sources all appeared as point-source in the ground-based data, however we found that in some cases the measured $r_{1/2}$ was larger than expected due to blending with foreground galaxies.

\begin{figure}
\begin{center}
\includegraphics[width =0.47\textwidth]{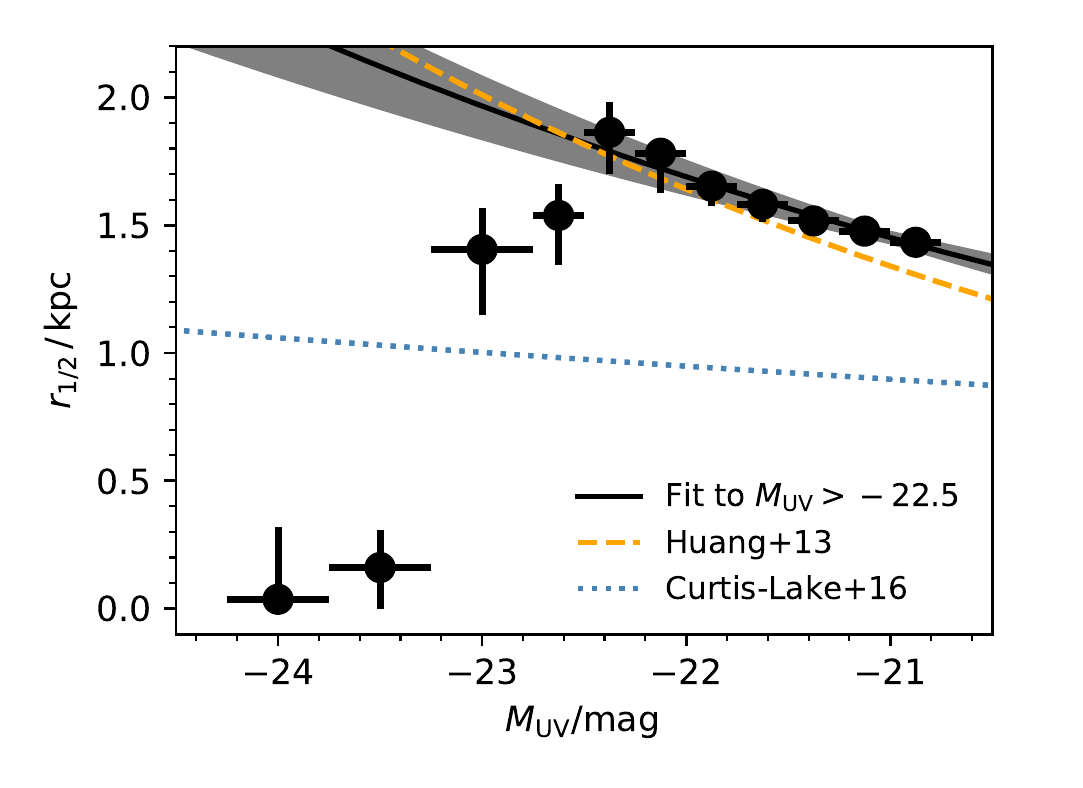}
\caption{The size-luminosity relation at $z \simeq 4$, derived from the subset of our sample that have high-resolution ACS $I_{814}$ coverage.
Our best-fit relation to the points at $M_{\rm UV} > -22.5$ is shown as the solid black line, with the grey shading showing the $1\sigma$ confidence interval.
The size-luminosity relation from~\citet{Huang2013} and~\citet{Curtis-Lake2016} at $z = 4$ are shown as the orange dashed and blue dotted lines respectively.
A clear deviation from the relation is observed at bright magnitudes, with the sources at $M_{\rm UV} < -23.2$ being consistent with being unresolved by~\emph{HST}/ACS.
}
\label{fig:slkpc}
\end{center}
\end{figure}

In order to measure the size-luminosity relation we corrected for the effect of the PSF by subtracting in quadrature the $r_{1/2}$ measured for stars in the imaging. 
This is an approximate correction for the PSF, however we use it for comparison with previous studies.
The size-luminosity relation was measured from the~\emph{HST}/ACS data only, as these data provide the most robust size measurements due to the smaller effect of the PSF (which has $r_{1/2}= 0.0725 \,{\rm arcsec}$).
We then converted the $r_{1/2}$ into a physical distance by assuming a redshift of $z = 4.0$.
The resulting size-luminosity relation derived from the~\emph{HST}/ACS data is shown in Fig.~\ref{fig:slkpc}.
As expected from this effective rescaling of the observed sizes shown in Fig.~\ref{fig:sl}, we see a clear drop in size at $M_{\rm UV } <-22.5$.
To determine the size-luminosity relation from the sample we therefore fit to the points faint-ward of this magnitude, assuming the standard parameterisation of $r_{\rm 1/2} \propto L^{\beta}$ (e.g.~\citealp{Shen2003}).
We find a best fitting slope of $\beta = 0.16 \pm 0.03$, with a normalisation given by $R_{0} = 1.45 \pm 0.02$ at $M_{\rm UV} = -21.0$.
If we use the {\sc SE} single-component centroid (e.g. prior to recombining de-blended components), we derive an identical slope, but a lower normalisation of $R_{0} = 1.36 \pm 0.02$.
We find no difference in results when using an average stack instead of the median presented here.
Our derived size-luminosity relation is consistent with that determined by~\citet{Huang2013}, who found $\beta = 0.22 \pm 0.06$, and~\citealp{Curtis-Lake2016} who found $\beta = 0.06 \pm 0.11$. 
We find an offset compared to the relation derived in~\citet{Curtis-Lake2016}, which we attribute to the different measurement of size used by that study.

By extrapolating our fitted size-luminosity relation to brighter magnitudes, it is evident that there is a dramatic drop in the observed sizes of $z \simeq 4$ sources.
We attribute this drop to the increasing contribution of point-sources bright-ward of $M_{\rm UV} = -22.5$, in agreement with the visual morphologies shown in Fig.~\ref{fig:acs}.
The transition occurs over almost one magnitude, being complete around $M_{\rm UV} = -23.25$ according to our ACS stacks.
From the XMM-LSS results shown in Fig.~\ref{fig:sl}, which cover a wider area than COSMOS and hence are likely to detect the presence of the rarest SF galaxies, there is evidence for extended sources up to $M_{\rm UV} \simeq -24$.

\subsection{Rest-frame UV spectroscopy}\label{sect:spectra}

\begin{figure*}
\begin{center}
\includegraphics[width =0.45\textwidth]{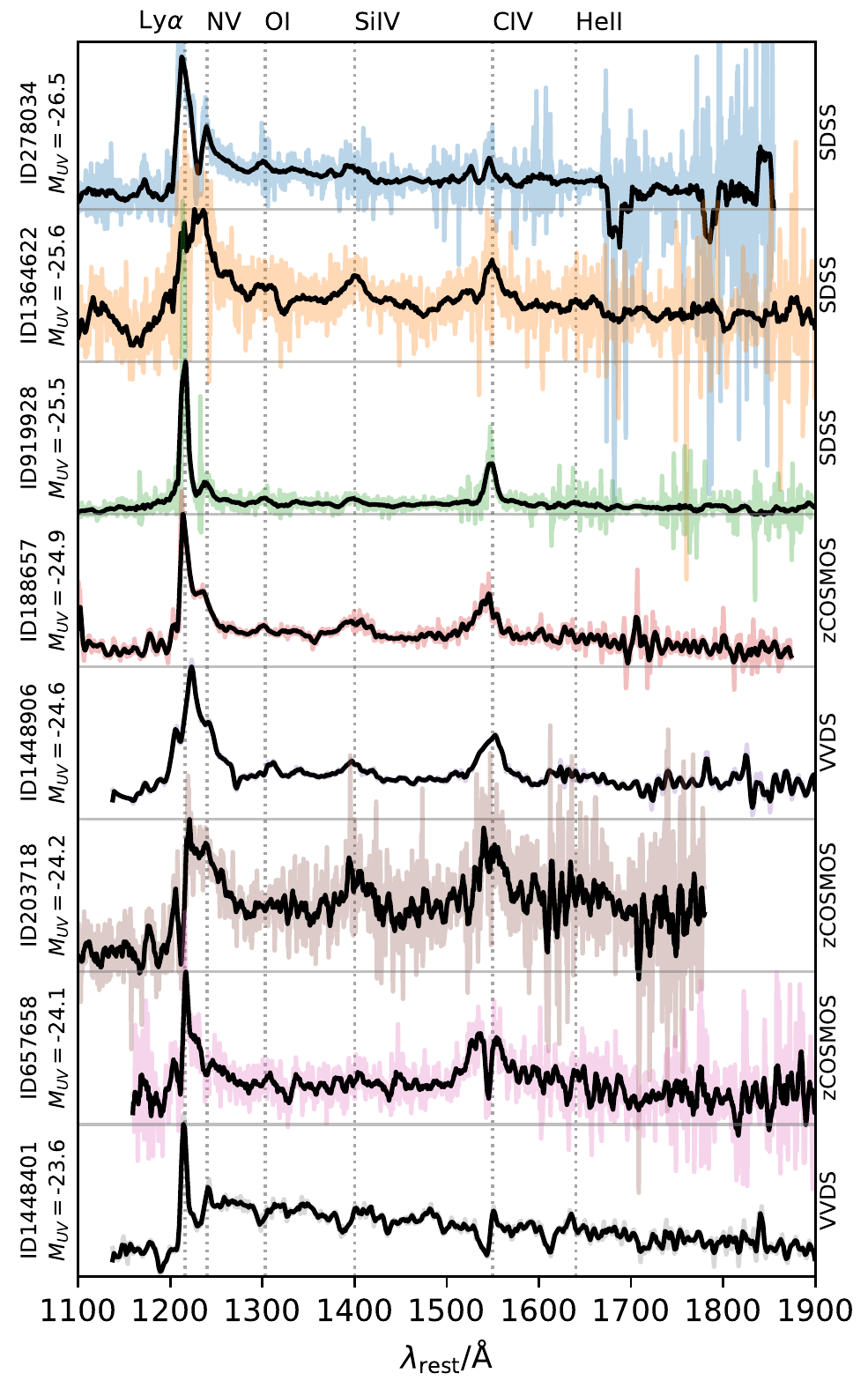}
\includegraphics[width =0.45\textwidth]{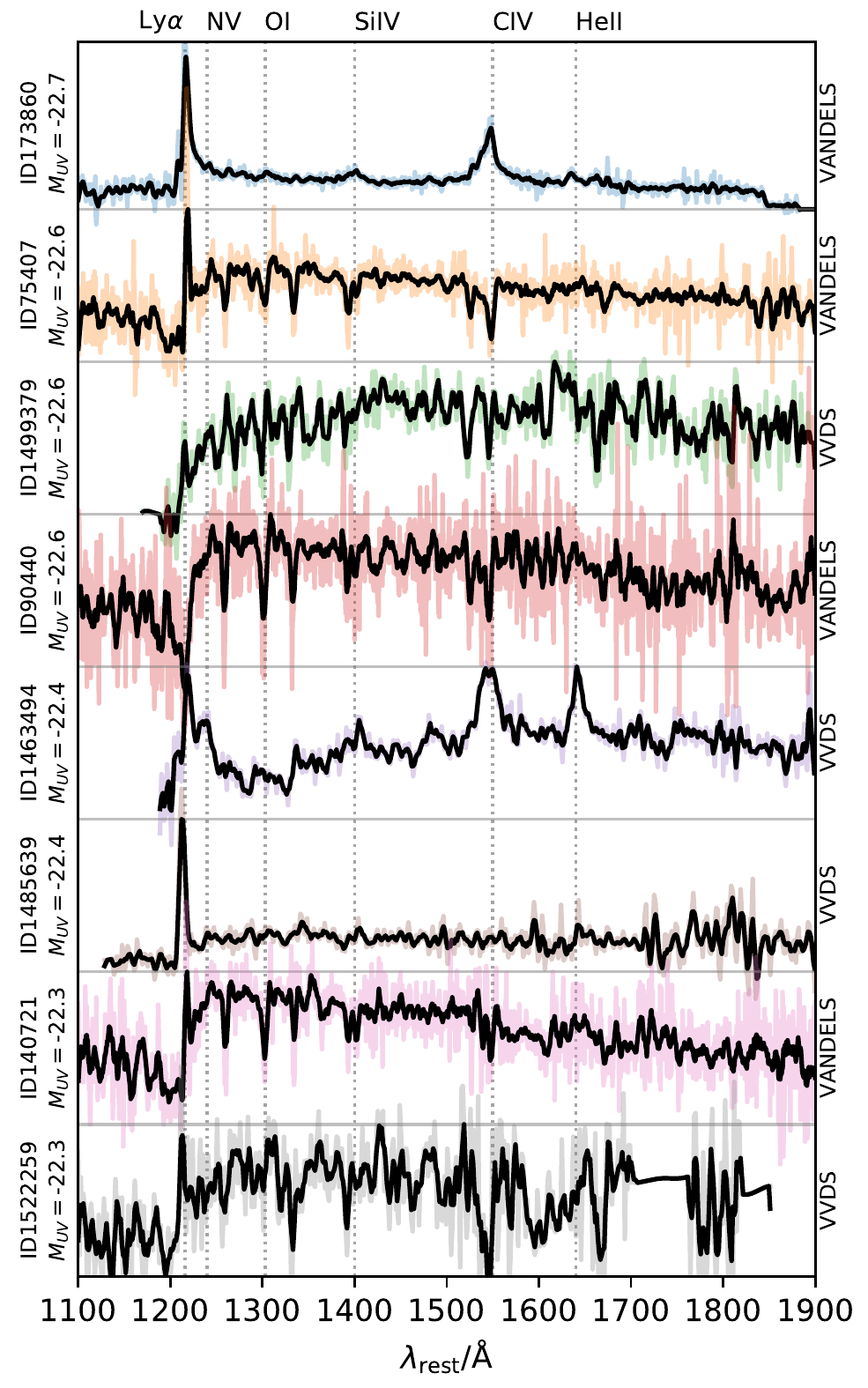}

\caption{A compilation of rest-frame UV spectra of the brightest sources in the $z \simeq 4$ sample.
The spectra have been shifted into the rest-frame according to the spectroscopic redshift provided by each survey, and are ordered by absolute UV magnitude with the brightest source at the top left.
The raw data is shown as the coloured background and a box-car filtered spectrum in shown in black.
Each spectrum (which was originally in units of ${\rm erg}/{\rm s}/$\AA) was normalised to a peak flux of 1.0, and has been presented offset in the vertical direction for clarity.
On the left of each spectrum is a label presenting the ID number and the absolute UV magnitude of the object.
On the right of each spectrum is the name of the survey which obtained the data.
We label common high-ionization emission lines with vertical dashed lines.
}
\label{fig:spectra}
\end{center}
\end{figure*}

\begin{figure}
\begin{center}
\includegraphics[width =0.47\textwidth]{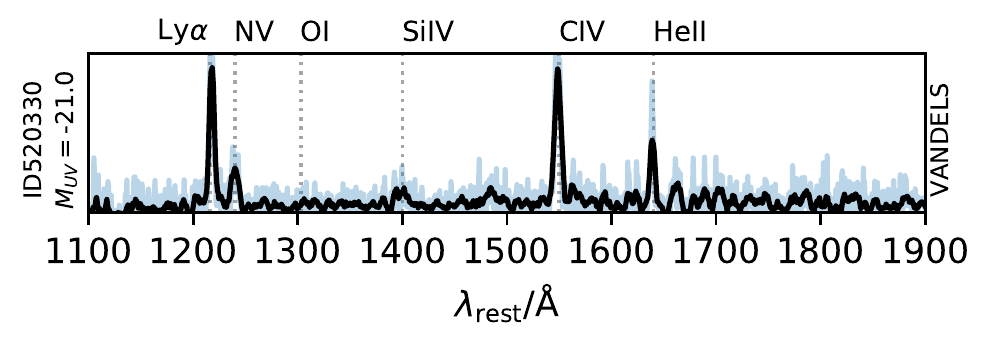}
\caption{The spectrum of the faintest AGN we have identified in our $z \simeq 4$ sample when cross-matched to publicly available spectra.
The data is displayed as in Fig.~\ref{fig:spectra}.
This source was found within the XMM-LSS field and has a spectrum from the VANDELS survey. 
It is the only source we find with AGN features faintward of $M_{\rm UV} = -22.4$.
}
\label{fig:faintspectra}
\end{center}
\end{figure}

To inform further our classification of sources as SF or AGN-dominated in the `transition' region observed in the size-luminosity relation, we retrieved the publicly available spectra available for the sample.
While the brightest sources in our sample are confirmed from magnitude limited surveys (e.g. SDSS, zCOSMOS), there is a dearth of spectra in the range $-23.5 < M_{\rm UV} < -22.5$, as is visible in Fig.~\ref{fig:hist} and shown in Table~\ref{table:spec}.
Nevertheless, we compiled the publicly available spectra and present the results for all sources brighter than $M_{\rm UV} = -22$ in Fig.~\ref{fig:spectra}. 
We smoothed the spectra with a box-car filter of width $1000\,{\rm km}/{\rm s}$ in the rest-frame, to highlight the spectral features above the noise.
The brightest seven sources show broad emission lines of {\sc NV} $\lambda 1240$\AA, {\sc SiIV} $\lambda \lambda 1393, 1402$ and  {\sc CIV}$\lambda \lambda 1548, 1550$\AA~in addition to strong Lyman-$\alpha$ emission, all of which are clear signatures of unobscured AGN spectra.
In addition to the typical AGN spectra, we identify source that show the appearance of SF-dominated light in the rest-frame UV.
Faint-ward of $M_{\rm UV} = -24$ we see a majority of sources that show the appearance of SF-dominated light, with narrow Lyman-$\alpha$ emission and absorption lines.
Of particular interest is ID1448401 which is the most luminous source ($M_{\rm UV} = -23.6$) in this sub-sample to show a SF-dominated spectrum.
We discuss the implications for the discovery of this object for the LF in Section~\ref{sect:lf}.
Within the SF-dominated objects, we see a large variation in the observed spectra, with some showing strong Lyman-$\alpha$ and others showing a continuum break and no appreciable Lyman-$\alpha$ emission.

The sources shown in Fig.~\ref{fig:spectra} are particularly bright which makes it relatively straightforward to see the presence of SF or AGN-type features in the spectra.
Even with this limited spectroscopic sub-sample, it is evident that there is a transition in the rest-frame UV spectra of $z \simeq 4$ sources between $-24.0 < M_{\rm UV} < -22.0$.
Fainter than $M_{\rm UV} = -22.4$ we find only one other spectrum that shows clear evidence of AGN signatures, both through a visual inspection of the smoothed data and through an analysis of the spectral flags provided by each survey.
The spectrum of this faint AGN is shown in Fig~\ref{fig:faintspectra}.
This source was observed as part of the VANDELS survey of the XMM-LSS field and has $z_{\rm spec} = 3.9407$ and $M_{\rm UV} = -21.0$.
It lies outside the region of~\emph{HST} data from CANDELS, however it appears extended in the ground-based HSC I-band data, with a PSF uncorrected $r_{\rm 1/2} = 0.77\,{\rm arcsec}$.
In the spectrum there are again strong emission lines of high-ionization species including {\sc CIV} and He{\sc II}.
In comparison to the brighter AGN shown in Fig.~\ref{fig:spectra} however, ID520330 shows stronger and considerably narrower emission lines.
We measured the FWHM of the {\sc CIV} doublet in all of the spectra shown, both directly from the smoothed data and through fitting a simple model of two Gaussians at the doublet wavelengths (constrained to have the same normalisation and standard-deviation).
Both methods produced consistent results within the errors.
The result of this analysis was that the brighter AGN in our sample show {\sc CIV} $FWHM \simeq 2000$--$6000\,{\rm km}/{\rm s}$, as found for the general population of SDSS quasars for example (e.g.~\citealp{VandenBerk2001}).
In contrast, the faint source ID520330 shows a significantly narrower width of $FWHM = 1200 \pm 100\,{\rm km}/{\rm s}$.
Similarly, this faint source has the highest rest-frame equivalent width of {\sc CIV} amongst the AGN spectra, showing $EW_{0} \simeq 150 \pm 30$\AA.
This is to be compared with $EW_{0} \simeq 20$--$90$\AA~for the brighter sources.
Emission lines of this width and strength are characteristic of obscured Type II AGN (e.g.~\citealp{Alexandroff2013}), where only the narrow line region is observed.
The fact that we see Type II signatures in the faintest source in the rest-frame UV is also to be expected, as the bright continuum from the AGN is obscured in this case.
Thus for this source we are observing predominantly the host galaxy continuum, with the addition of AGN emission lines.

\section{Separating the UV LF of AGN and LBGs}\label{sect:lf}

As expected, the brightest sources in our $z \simeq 4$ sample appear to be AGN-dominated in the rest-frame UV, showing a point-source morphology in the~\emph{HST}/ACS imaging and strong quasar features in the rest-frame UV spectra.
Faint-ward of $M_{\rm UV} \simeq -22.5$ however, the sources become extended, with spectra that are dominated by the light from young stars.
In the `transition' regime between these AGN and SF-dominated objects, we find evidence for a mixture of these two classes in our sample.
In this Section we use these observations to infer the rest-frame UV LF of the two components, with the assumption that the majority of the sources in our sample can be separated into either an AGN or SF-dominated category.
The existence of a slightly extended source that shows the rest-frame UV spectrum of an AGN (see Section~\ref{sect:vis}) demonstrates that this assumption will break down depending on how AGN are distributed in the underlying galaxy population.
We discuss this issue further in Section~\ref{sect:model}.

\subsection{AGN fraction}\label{sect:agnfrac}

To separate AGN and galaxies in our sample we define an AGN fraction ($f_{\rm AGN}$) as a function of absolute UV magnitude.
We first define a quantitative measure of AGN-dominated sources from the $I_{814}$-band high-resolution images by assigning objects with a $r_{1/2} < 0.1\,{\rm arcsec}$ as AGN.
In addition, we determined a comparison $f_{\rm AGN}$ from the visual morphology of the brightest sources.
Despite this being more subjective than a size cut we found very close agreement between these two measures.
As a final check we also used {\sc GALFIT} to fit the stacked images in each $M_{\rm UV}$ bin.
We used a two component model consisting of a point-source and a S{\'e}rsic profile (with a fixed index of $n = 1$). 
Reassuringly the AGN fraction of each stack, as defined by the ratio of the flux in the point-source compared to the total flux, agreed very well with the size cut.
Hence we are confident that the derived AGN fraction from the source morphology does not depend significantly on the method used to derive it.
Faint-ward of $M_{\rm UV} = -22$, the smaller mean sizes of galaxies coupled with the scatter in the galaxy size-luminosity relation makes it more challenging to separate compact galaxies from point-sources using a size criterion.
Hence we do not present measurements of the morphology-based $f_{\rm AGN}$ for sources faintward of $M_{\rm UV} = -22$.
We also defined an AGN fraction from archival spectra for a sub-set of our sample.
Using the spectra presented in Fig.~\ref{fig:spectra} we identified AGN-dominated sources according to the presence of strong emission lines of {\sc CIV}, {\sc NV} and He{\sc II}.
Due to the small number of sources with spectra we used wider bins than for our morphology measurement.
We determined the centre of each bin by taking the average source luminosity to negate any bias in the distribution of sources within that bin.
Faintward of $-22.0$ we find only one source which has AGN signatures in the available spectroscopy sample.
This source is identified as a Type II AGN (ID520330) in which the rest-frame UV continuum is dominated by the host-galaxy light and we therefore define this source as SF-dominated (as discussed in Section~\ref{sect:typeII}).

\begin{figure}
\begin{center}
\includegraphics[width =0.5\textwidth]{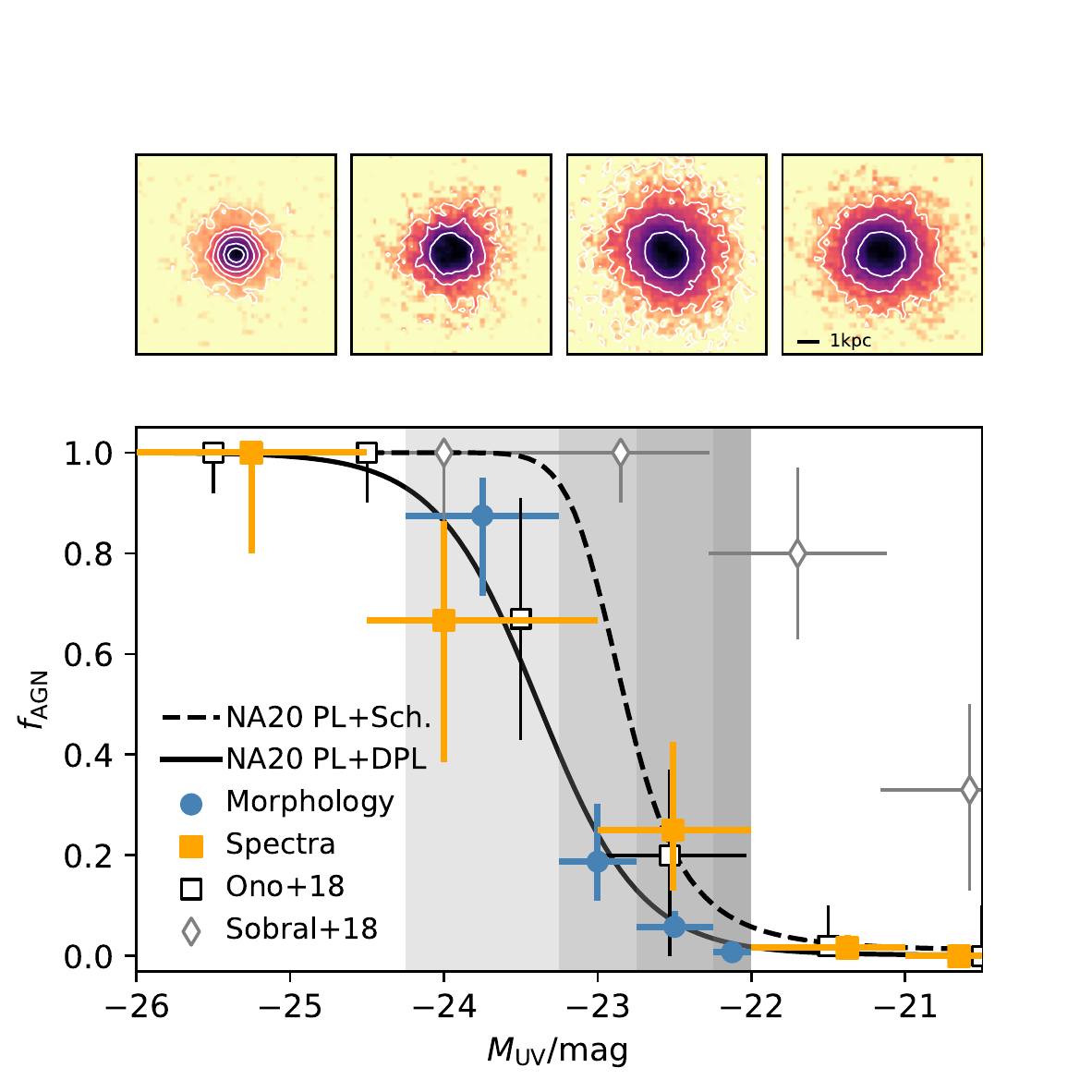}
\caption{
The AGN fraction as a function of absolute UV magnitude at $z = 4$, derived from morphology/size criteria (blue circles) and from spectroscopy (orange squares).
We compare to previous estimates of the AGN fraction at $z = 4$ from~\citet{Ono2018} and the $z =2$--$3$ results from~\citet{Sobral2018} as the open black squares and grey diamonds respectively.
The lines show the predicted $f_{\rm AGN}$ from the simultaneous fitting of the LBG and AGN LF presented in~\citet{Adams2020}.
The dashed and solid lines show the results assuming a Schechter and DPL form for the LBG LF respectively.
The four images in the upper row show the result of stacking our sample in the four grey highlighted bins in $M_{\rm UV}$ shown on the plot.
The stamps are $1.5\,{\rm arcsec}$ on a side, with contours at intervals of $1.0\,{\rm mag}$ from the peak.
}
\label{fig:agnfrac}
\end{center}
\end{figure}

We present the derived $f_{\rm AGN}$ measurements in Fig.~\ref{fig:agnfrac}.
Both the morphological and spectroscopic measurements show a sharp drop in the $f_{\rm AGN}$ between $-24 \lesssim M_{\rm UV} \lesssim -22.5$, with around equal occurrence of AGN and LBGs at around $M_{\rm UV} \simeq -23.2$.
This is also visually apparent in the stacked~\emph{HST}/ACS images (top of Fig.~\ref{fig:agnfrac}), where at $M_{\rm UV} \simeq -23$ the stack is clearly extended (although with less flux in the wings) while at $M_{\rm UV} \simeq -23.8$ the image is consistent with being a point-source.
Comparing to previous estimates of the $f_{\rm AGN}$ at $z \simeq 4$ we find good agreement with the spectroscopy measurements of~\citet{Ono2018}, who used predominantly archival redshifts with spectroscopic flags to determine the AGN fraction.
The advantage of our method of AGN classification from the full spectrum is that we are not sensitive to differences between AGN classifications between spectroscopic surveys.
We find a brighter transition magnitude than that of~\citet{Sobral2018}, who found a drop in the fraction of AGN at $M_{\rm UV} = -21.5$ at $z \simeq 2$--$3$.
This study was based on the follow-up of strong Lyman-$\alpha$ emitters rather than LBGs, and hence it could be expected that this pre-selection for strong line-emitters would preferentially detect AGN at fainter magnitudes.
We note however, that in~\citet{Sobral2018} the AGN fraction at $M_{\rm UV} > -22.5$ is determined from sources predominantly from a detection of the {\sc NV} line with low S/N ($\lesssim 3.0$), their classification as AGN is somewhat uncertain.

At $-23 < M_{\rm UV} < -22$ we see a slight difference in the derived AGN fraction from our morphological and spectroscopic measurements.
From a morphology cut we measure $f_{\rm AGN} = 0.06^{+0.03}_{-0.02}$ at $M_{\rm UV} = -22.5$ while with spectroscopic data we find $f_{\rm AGN} = 0.25^{+0.17}_{-0.12}$.
Although the errors are large, due predominantly to the small number statistics for the spectroscopic sub-sample, a difference between the $f_{\rm AGN}$ between a strict morphological selection and spectroscopic identification could be expected in this magnitude range.
This is a consequence of the increasing importance of host galaxy light at fainter UV magnitudes, and we present a toy model that can explain these observations in Section~\ref{sect:model}.
Alternatively the slight difference found could be due to bias in the spectroscopic measurement, as arguably the strong emission lines from AGN and the compactness of the emission could make them easier to identify in spectroscopic measurements.
In the range $-23 < M_{\rm UV} < -22$ we find no difference between the sizes of the spectroscopically confirmed sources and our full sample, suggesting that we are not biased to compact sources.
In this magnitude range we find 10 sources with archival spectroscopy, four from the VANDELS survey and six from VVDS (of which eight with secure flags are shown in Fig.~\ref{fig:spectra}).
While VVDS is a purely I-band magnitude limited survey, VANDELS selected against compact sources over 50 percent of the survey area that was covered by~\emph{HST} imaging~\citep{McLure2018}, and hence VANDELS should be biased~\emph{against} AGN.
If we measure the $f_{\rm AGN}$ in the VVDS and VANDELS survey separately, we find $f_{\rm AGN} = 0.25^{+0.25}_{-0.15}$ for both surveys when secure flags are used, indicating that there is no clear bias within the limitations of small number statistics.
Two of the VVDS spectra were not included in our initial AGN fraction calculation as they have poor quality flags.
If we assume these objects are SF-dominated, under the assumption that AGN features are easier to identify, then we obtain a lower $f_{\rm AGN} = 0.20^{+0.15}_{-0.10}$ in this magnitude range. 
This value is still higher than that derived from our morphology measurement, however larger spectroscopic samples are clearly required to determine if there is a real discrepancy between the $f_{\rm AGN}$ found from a strict morphological cut in contrast to a classification from spectroscopy.

In Fig.~\ref{fig:agnfrac} we also present the predicted AGN fraction from the simultaneous fitting of the combined AGN and LBG LF presented in~\citet{Adams2020}.
\citet{Adams2020} assumed either a Schechter function or DPL form for the LBG LF in addition to a single power-law (PL) to model the faint-end of the AGN.
Without any further information about the nature of the sources, both models produced a good fit for the observed UV LF over $-26 < M_{\rm UV} < -20$ (see left panel of Fig.~\ref{fig:lf}).
In comparison to our derived $f_{\rm AGN}$ we see that the Schechter function form of the LBG LF predicts a steeper decline in the fraction of AGN at fainter magnitudes than a DPL model, due to the exponential drop-off at the bright-end of this parameterisation.
Note that the position of this drop depends on the position of the `knee' in the LF, which is strongly constrained by the number density of sources $M_{\rm UV} > -21$.
Both our measurements of the AGN fraction deviate from this steeper Schechter prediction at $M_{\rm UV} \sim -24$, as do the results of~\citet{Ono2018}, suggesting that a DPL is the more appropriate function to describe the LBG LF at $z \simeq 4$.

\subsection{The Luminosity Function}

\begin{figure*}
\begin{center}
\includegraphics[width =\textwidth]{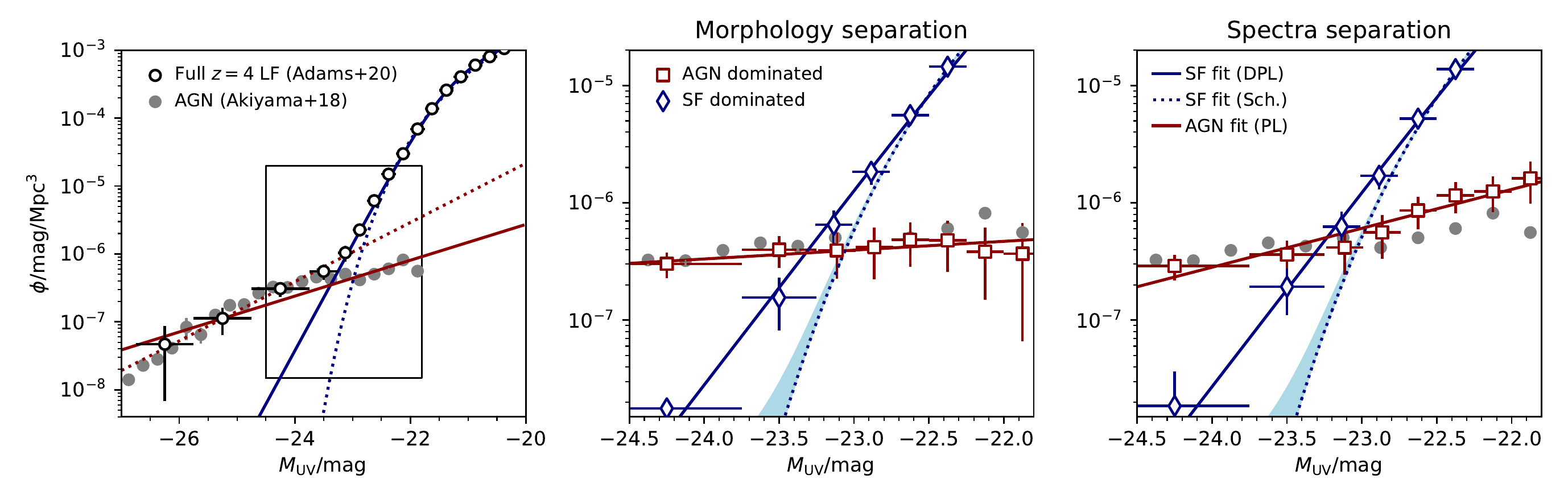}
\caption{The rest-frame UV LF at $z \simeq 4$.
The left-hand panel shows the full LF derived in~\citet{Adams2020} as the open black circles.
The solid (dotted) lines on this plot show the result of the simultaneous fitting presented in~\citet{Adams2020} with a DPL (Schecher) assumed galaxy LF.
The central and right-hand plots show a zoom-in of the transition region, where we have separated objects that have rest-frame UV light that is SF (blue diamonds) or AGN (red squares) dominated.
In these plots we show the best-fit AGN power law as the red line.
The best-fit Schechter or DPL function to the SF-dominated results are shown as the blue dotted and solid lines respectively.
For the Schechter function the fit is constrained by points that are fainter than $M_{\rm UV} = -22$.
The effect of the magnification bias on the Schechter function is shown as the blue shaded excess on this curve.
In all three plots we show the AGN results from~\citet{Akiyama2018} as the grey filled circles.
}
\label{fig:lf}
\end{center}
\end{figure*}

In the previous section we compared the observed $f_{\rm AGN}$ to that expected from the fitting of the full AGN + LBG rest-frame UV LF.
In this section we instead use the $f_{\rm AGN}$ derived in this study to separate the LF results of~\citet{Adams2020} into AGN and SF-dominated subsamples.
Because we could only classify a small fraction of the full sample using the morphology and spectroscopy data (e.g. because high-resolution imaging was only available in COSMOS, and the spectroscopy data only covers $\sim 1$ percent of the sample), we elected to apply the $f_{\rm AGN}$ to the data points from~\citet{Adams2020} as opposed to recalculating the LF from a significantly smaller sample.
We determined the separate LFs by using the AGN fractions derived from the morphology and spectroscopy results separately.
To interpolate the $f_{\rm AGN}$ we fit a constrained model to our binned AGN fraction points shown in Fig.~\ref{fig:agnfrac}.
The model consisted of two power-laws to approximate the overlap between the bright-end of the galaxy LF and the faint-end of the AGN LF without over-fitting the data.
Due to the difference in the $f_{\rm AGN}$ derived from using the morphological and spectroscopic data, we find differences in the separate LFs for the AGN and SF dominated sources as shown in the two right-hand plots in Fig.~\ref{fig:lf}.

We present the results of fitting the separated AGN and SF-dominated LFs with different parameterisations in Fig.~\ref{fig:lf}.
The best-fit parameters are presented in Table~\ref{table:lf} in comparison to the LF parameters derived from the simultaneous fit of~\citet{Adams2020}.
The AGN fractions derived from these fits in comparison to the results of~\citet{Adams2020} are presented in Appendix~\ref{sect:appendixB}.
We fit our separated AGN-dominated UV LFs using a single power-law, as we do not extend faint-ward of the apparent LF knee at $M_{\rm UV} \sim -26$.
For the SF LF we fit using both a Schechter function or a DPL.
If we focus first on the SF-dominated results, we find that the separation of sources using a morphology or spectroscopy criterion makes only a marginal difference to the derived LF. 
This is evident in the fitting results presented in Table~\ref{table:lf}, where the values are well within the $1\sigma$ errors in the different scenarios.
We also find good consistency with the Schechter and DPL parameters from~\citet{Adams2020}, which is to be expected as the LBG fit is predominantly constrained by the data-points at $M_{\rm UV} > -22$.
We checked that the impact of strong gravitational lensing on a Schechter function fit could not reproduce the number of bright sources by applying the methodology of~\citet{Mason2015, Barone-Nugent2015a}.
The excess that results from the lensing is shown in Fig.~\ref{fig:lf} and is too small to account for the number density we find at $M_{\rm UV} < -23$.

In contrast to the SF-dominated LF, the faint-end of the AGN-dominated LF depends more significantly on the assumed $f_{\rm AGN}$.
If we use a morphology criterion we find a shallow slope ($\alpha = -1.19 \pm 0.05$) due to the rapid drop in point-source dominated sources faint-ward of $M_{\rm UV} \simeq -23$.
In this case, we find close agreement with the results of~\citet{Akiyama2018}, who derived a faint-end slope of $\alpha = -1.30 \pm 0.05$.
This is to be expected given that~\citet{Akiyama2018} identified AGN based on a compactness criterion in the ground-based HSC data.
If instead we use our spectroscopic criterion in determining $f_{\rm AGN}$ we find a significantly steeper slope of the faint-end of the AGN LF of $\alpha = -1.85 \pm 0.05$.
In this case our data points start to diverge from the~\citet{Akiyama2018} points, due to a higher proportion of AGN-dominated sources at faint magnitudes in this parameterisation (see Fig.~\ref{fig:agnfrac}).
\citet{Adams2020} found $\alpha = -1.66^{+0.29}_{-0.58}$ in the fit of a DPL (LBG component) with a PL (AGN component).
The large error on this value is a consequence of the degeneracy between the bright-end slope of the LBG LF and the AGN faint-end slope.
Our measurement of the slope of the faint-end of the AGN LF is more constrained by the data-points at $M_{\rm UV} \gtrsim -24$, which allows us to find a best-fit that is shallower, but still consistent within $1.5\sigma$, from the~\citet{Adams2020} result.

\begin{table}
\caption{The luminosity function parameterisations for the separated SF and AGN results shown in Fig.~\ref{fig:lf}.
The upper part of the table shows the results when separating according to a morphological criterion, while the lower part shows the results when AGN are identified according to their spectra.
The first column indicates the sub-sample that was fit to (SF-dominated or AGN-dominated).
For the SF case, we show the results for a Schechter function and DPL fit in the first and second row.
The fit to the AGN case was performed with a single power law, with the normalisation calculated at a fixed $M_{\rm UV}$ highlighted with an asterick.
The second and third column denote the characteristic absolute magnitude and normalisation.
The fourth column show the faint-end slope for the SF and AGN fits, and the final column shows the bright-end slope for the DPL fit.
}
\setlength\tabcolsep{4pt}
\def\arraystretch{1.15}
\begin{tabular}{lcccc}
\hline
Type & $M^*$ & $\phi^*$ & $\alpha$ & $\beta$ \\
& $/{\rm mag}$ & $/{\rm mag}/{\rm Mpc}^3$ & & \\
\hline
SF & $-21.00(0.10)$ & $1.36(0.24) \times 10^{-3}$ & $-1.75(0.13)$ & -- \\
SF & $-21.53(0.06)$ & $0.36(0.05) \times 10^{-3}$ & $-2.07(0.07)$ & $-5.15(0.10)$ \\
AGN & -25.70* & $2.48(0.31) \times 10^{-7}$ & $-1.19(0.05)$ & -- \\
\hline
SF & $-20.97(0.09)$ & $1.44(0.25) \times 10^{-3}$ & $-1.72(0.13)$ & -- \\
SF & $-21.50(0.05)$ & $0.37(0.04) \times 10^{-3}$ & $-2.05(0.06)$ & $-5.15(0.09)$ \\
AGN & -25.70* & $0.77(0.21) \times 10^{-7}$ & $-1.83(0.11)$ & -- \\
\hline
\multicolumn{5}{c}{\citet{Adams2020}}\\
\hline
Sch. & $-20.89^{+0.12}_{-0.10}$ & $1.62^{+0.33}_{-0.27} \times 10^{-3}$ & $-1.66^{+0.13}_{-0.08}$ & -- \\
+PL & -25.70* & $ 0.71^{+0.44}_{-0.39} \times 10^{-7}$ & $-2.09^{+0.32}_{-0.38}$ & -- \\
\hline
DPL & $-21.37^{+0.08}_{-0.11}$ & $ 0.50^{+0.10}_{-0.06}\times 10^{-3}$ & $-1.92^{+0.07}_{-0.04}$ & $-4.92^{+0.29}_{-0.25}$ \\
+PL & -25.70* & $0.85^{+0.81}_{-0.34} \times 10^{-7}$ & $-1.66^{+0.29}_{-0.58}$ & -- \\

\hline
\end{tabular}
\label{table:lf}
\end{table}

\section{Discussion}\label{sect:dis}

In this work we have investigated the transition in the properties of $z \simeq 4$ sources at $M_{\rm UV} \simeq -23$, where the number densities of faint-AGN and bright-galaxies converge.
From our imaging data we observe a change in the source morphology, through a sharp drop in the average size of sources in the size-luminosity relation that is also seen in the individual source morphology.
We also see a change in the features present in the available rest-frame UV spectra for the sample.
The absolute UV magnitude at which this transition occurs corresponds to the point of rapid decline in the bright-end of the galaxy LF.
Furthermore, the form of the increase in the AGN fraction to brighter magnitudes depends on the shape of the galaxy LF bright-ward of the knee in the function ($M_{\rm UV} \simeq -21$; see Table~\ref{table:lf}).
There has been an ongoing discussion on the shape of the rest-frame UV at high-redshifts.
While the UV LF at $z \gtrsim 4$ has typically been fitted by a Schechter function (e.g.~\citealp{McLure2013, Finkelstein2015, Bouwens2015}), recent results have demonstrated an excess of highly luminous galaxies in relation to the Schechter function predictions~\citep{Bowler2014, Ono2018, Bowler2020}.
In our derived AGN fraction, and in the corresponding SF-dominated LFs, we have found evidence for a shallower decline in the number density of the brightest SF galaxies at $z \simeq 4$.
Most strikingly, the discovery of an extremely bright source at $M_{\rm UV} = -23.6$ with no evidence for AGN spectral features (ID1448401 in Fig.~\ref{fig:spectra}) supports a $f_{\rm AGN} \simeq 0.8$ (range within the errors of $0.38$--$0.95$) at this magnitude, which leads to a number density of sources well in excess of the Schechter function prediction.
This finding is potentially in conflict with studies that have found support for a Schechter function form at $z \simeq 3$--$4$~\citep{VanderBurg2010, Hathi2010, Bian2013, Parsa2016}, however it is only recently that the datasets available at $z \simeq 4$ have had sufficient volume to adequately constrain the number density of the rarest galaxies/faint-AGN.
If we fit our SF-dominated LF with a DPL, the results are in good agreement with the evolution in the DPL parameters derived in~\citet{Bowler2015, Bowler2020}, who found a steady steepening of the bright-end from $z \simeq 9$ to $z \simeq 5$ consistent with the increasing impact of dust.
If this steepening is due to dust obscuration in the most highly star-forming galaxies, then the effects of this dust should be observable both in the colours of bright LBGs and directly via reprocessed emission in the far-infrared.
Interestingly, the brightest spectroscopically confirmed LBG in our sample (ID1448401) shows a very blue rest-frame UV continuum (rest-frame UV slope $F_{\lambda} \propto \lambda^{\beta}$; $\beta \simeq -2$).
From the observed colour-magnitude relation at this redshift, this source would be expected to show a redder slope with $\beta \simeq -1.4$~\citep{Lee2011, Bouwens2014beta}.
\citet{Rogers2014} demonstrated that at $z \simeq 5$ there is an increased scatter in the rest-frame UV slopes of LBGs to brighter magnitudes, and thus it is plausible that this LBG is a rare example of a highly star-forming galaxy ($SFR \simeq 80\,{\rm M}_{\odot}/{\rm yr}$;~
\citealp{Madau1998}) with little dust attenuation at this redshift.
Thus while overall the increased production and attenuation of dust in the most highly SF galaxies from $z \simeq 9$--$4$ could cause a steepening of the bright-end slope of the rest-frame UV LF, this does not preclude the existence of galaxies with high-SFRs and a lack of dust obscuration within this epoch.

\subsection{The faint-end of the AGN UV LF}

There has been a renewed interest in recent years on the slope of the faint-end of the rest-frame UV LF of AGN.
This was motivated by the claimed detection of high-redshift X-ray sources by~\citet{Giallongo2015}, who used their data to suggest that UV-faint AGN could contribute significantly to the process of reionization at $z > 6$.
Such an analysis relies on the integral of an extrapolated rest-frame UV LF to determine the total number of ionizing photons that can be produced by AGN at very high-redshifts.
The subsequent studies of~\citet{Boutsia2018} and~\citet{Giallongo2019} have further claimed an excess in sources at the faint-end of the $z \simeq 4$.
While several works have called these results into question (e.g.~\citealp{Parsa2018, McGreer2018, Cowie2020}) at $z > 4$, the determination of an accurate slope of the faint-end of the AGN LF remains of interest.
Our observations demonstrate that the derived slope of the $z \simeq 4$ AGN LF depends strongly on the selection method.
We can reproduce the flatter slope found in~\citet{Akiyama2018} by using a criterion on morphology to separate AGN-dominated sources from the full $z \simeq 4$ LF.
If instead we use a spectroscopic determination of $f_{\rm AGN}$ to estimate the AGN LF, we derive a steeper faint-end slope ($\alpha \sim -1.8$).
The rest-frame UV spectroscopic features of AGN are strong and broad emission lines (e.g. Fig.~\ref{fig:spectra}), while LBGs are expected to show absorption features or potentially weak nebular emission lines~\citep{Stark2014, Shapley2003, Steidel2016}.
Thus we expect any classification of a source as an AGN based on the rest-frame UV spectrum to be sensitive to not only the brightest AGN-dominated objects, but also objects in which the light from SF is significant.
Instead, AGN selections that include a point-like morphology selection will exclude objects where the host galaxy UV light causes the source to be rejected as too extended.
Given the wide range in imaging depths and compactness criterion used in different AGN selections, it is challenging to understand the incompleteness of previous studies due to this effect.
It is clear however, from our study and other works (e.g.~\citealp{Matsuoka2018a}), that faint-ward of $M_{\rm UV} \simeq -24$ it is necessary to account for both AGN and SF-dominated sources.

\subsection{Evolution of the AGN to SF transition into the EoR}

Both the AGN and LBG rest-frame UV LFs are known to evolve rapidly at $z \gtrsim 4$.
In~\citet{Bowler2020} we found evidence for a flattening of the bright-end slope with increasing redshift in the range $z = 5$--$10$, with a corresponding evolution in $M^*$ according to $\Delta M^* /\Delta z \simeq -0.5$.
This was interpreted as a result of decreased dust obscuration and mass quenching within the Epoch of Reionization (EoR).
Such a change in shape would be imprinted onto the measured AGN fraction at these redshifts, with a predicted fainter transition magnitude and extended tail of highly luminous SF galaxies.
This prediction is consistent with the tentative detection of weak AGN features in the rest-frame UV spectra of moderately bright LBGs at $z \gtrsim 6$~\citep{Laporte2017, Tilvi2016}, as we predict a higher $f_{\rm AGN}$ to fainter magnitudes within this epoch.
These detections however, are at odds with the expected number density of faint AGNs at $z \simeq 6$, which have been shown to undergo an accelerated decline at $z > 5$~\citep{McGreer2013, Jiang2016}.
From the extrapolated number densities of high-redshift AGN, we do not expect that current LBG samples at $z \ge 7$ will contain any AGN-dominated sources, as AGN are only expected to be more numerous than LBGs at $M_{\rm UV} \le -24$ (see discussion in~\citealp{Bowler2014}).
This is consistent with the lack of point-sources found at $z \simeq 6$--$7$, where it has been possible to gain high-resolution imaging of the brightest sources with~\emph{HST}~\citep{Jiang2013b, Bowler2017}.
These somewhat conflicting results could be a result of weaker AGN residing within high-redshift LBGs, or misclassification of emission lines as AGN signatures.
Taking the results of this study coupled with what is known about the evolving UV LFs to higher redshifts, we predict that AGN `contamination' of rest-frame UV selection samples faint-ward of $M_{\rm UV} \simeq -23$ will be minimal at $z \ge 7$.

\section{A simple model of the AGN UV LF}\label{sect:model}

We created a toy model of the predicted rest-frame UV LF of AGN to aid in the interpretation of our observations.
The model takes the observed UV LF of LBGs and uses this, via simple empirical relations, to estimate the luminosity and number density of UV-bright AGN at the same epoch.

\subsection{Method}
For each galaxy of a given absolute UV magnitude, we first estimate the stellar mass according to the relation found by~\citet{Duncan2014}:

\begin{equation}
{\rm log}_{10}(M_{\star}) = (9.02 \pm 0.02)\, (M_{\rm UV} +19.5)  - (0.45 \pm 0.02).
\end{equation}

The slope and normalisation of this relation is consistent between different studies (e.g.~\citealp{Salmon2015, Song2016}; see figure 5 of~\citealp{Tacchella2018}).
This relation has been derived in the past to determine the stellar mass functions at high-redshift from the rest-frame UV LF, where the effect of scatter is essential in order to reproduce the observed mass and LFs~\citep{Stark2013, Duncan2014}.
We therefore include a scatter of $0.4\,{\rm dex}$ in the relationship above, which is consistent with that required in~\citet{Duncan2014} and is at the upper end of the measured intrinsic scatter in the $SFR$--$M_{\star}$ relation~\citep{CurtisLake2020, Salmon2015}.
From the stellar mass we then estimate the black hole mass using the $m_{\rm BH} - M_{\star}$ relation of the form:

\begin{equation}
{\rm log}_{10}(m_{\rm BH}) = G_{\rm BH} \, [{\rm log}_{10}(M_{\star}) +11.0]  + I_{\rm BH}
\end{equation}

Here $G_{\rm BH} = d{\rm log}_{10} (m_{\rm BH})/dM_{\star}$ gives the gradient of the relation and $I_{\rm BH}$ the intercept (defined at a mass of ${\rm log}_{10}(M_{\star}/{\rm M}_{\odot}) = 11.0$).
The form or even existence of such a relationship at high-redshift is uncertain.
We therefore consider two plausible scenarios based on previous results from both luminous quasars at high-redshift and low-redshift galaxies.
The simplest scenario is one in which the black-hole mass is a constant fraction of the stellar mass at a given redshift.
In this case (denoted model A hereafter) we set $G_{\rm BH} = 1.0$ and $I_{\rm BH} = 9.0+ {\rm log}_{10}(1+z)$ to give $m_{\rm BH}/M_{\star} = 0.05$ at $z = 4$, as found in observations of high-redshift quasars (e.g.~\citealp{Venemans2017, Targett2012}).
The $1+z$ term follows observations and theoretical arguments for an increased $m_{\rm BH}$ to bulge-mass ratio at high-redshifts (e.g.~\citealp{Venemans2015c, Croton2006, Wyithe2003}).
Such high ratios of $m_{\rm BH}/M_{\star}$ may not be representative of the AGN population at this redshift (e.g due to selection effects), hence we treat this scenario as an extreme case.
An alternative scenario is one in which black holes in more massive galaxies are over-developed, whereas those in less massive sources are a smaller fraction of the stellar mass.
Such a scenario has been measured at low-redshift by~\citet{Reines2015}.
In this case (denoted as model B hereafter) we set $G_{\rm BH} = 1.4$ and $I_{\rm BH} = 8.95 + {\rm log}_{10}(1+z)$.
Scatter is also significant in the $m_{\rm BH}$-$M_{\star}$ relation (e.g.see~\citealp{Hirschmann2010, Volonteri2016}), and we therefore include an intrinsic scatter of $0.3\,{\rm dex}$.
The result of these steps is a relationship between the $M_{\rm UV}$ of a LBG and the estimated $m_{\rm BH}$.
The $m_{\rm BH}$ can then be converted into an estimated bolometric luminosity using an assumption on the Eddington ratio.
We assume a log-normal distribution of mean $\lambda = 0.6$ and $\sigma = 0.3\,{\rm dex}$ as found by~\citealp{Willott2010a} (see also~\citealp{Kelly2013}).
Finally we convert the bolometric luminosity into a UV luminosity by assuming a bolometric correction (taken to be $4.4$;~\citealp{Runnoe2012,Mortlock2011}).
From a single $M_{\rm UV}$ from SF we thus obtain a spread in the predicted total absolute UV magnitude due to the addition of an unobscured black hole ($M_{\rm UV, BH}$).

\begin{figure}[h]
\begin{center}
\includegraphics[width =0.48\textwidth]{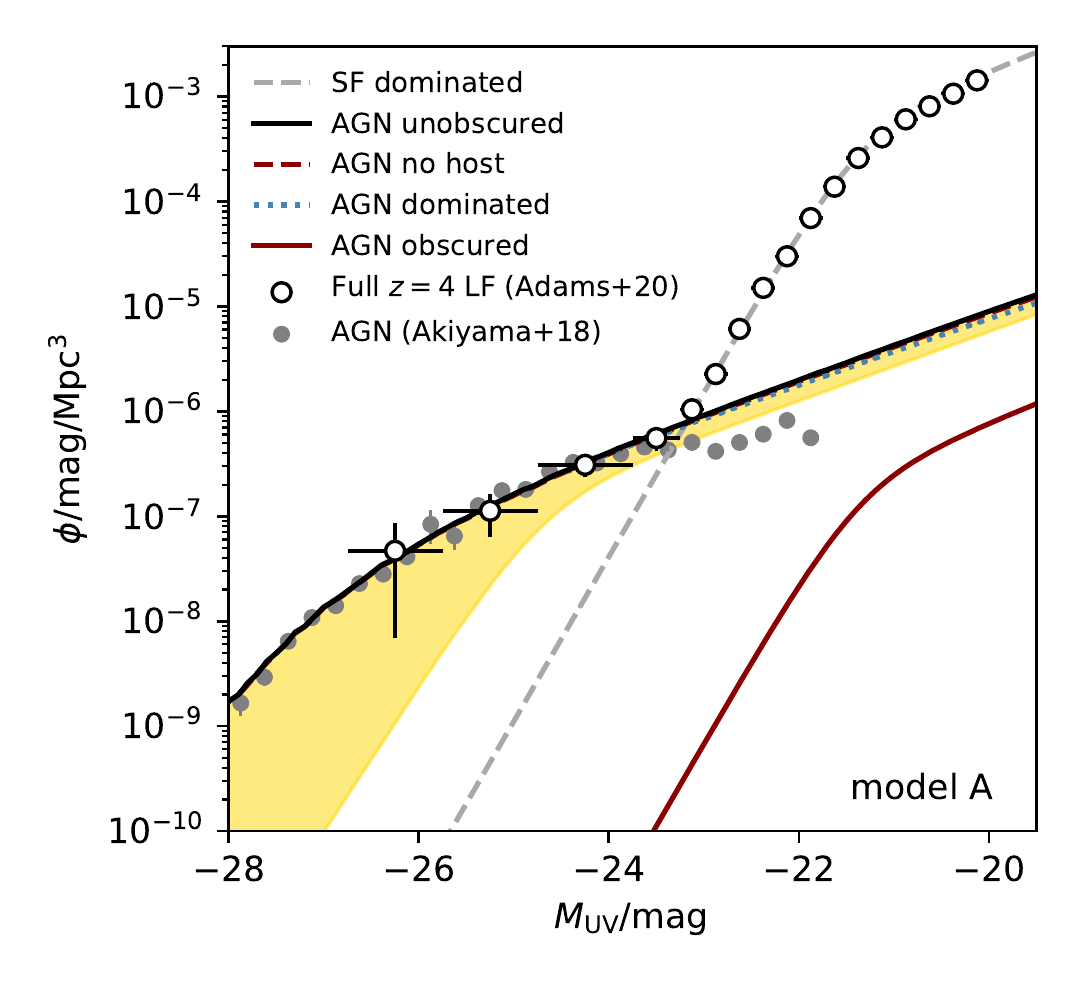}\\
\includegraphics[width =0.48\textwidth]{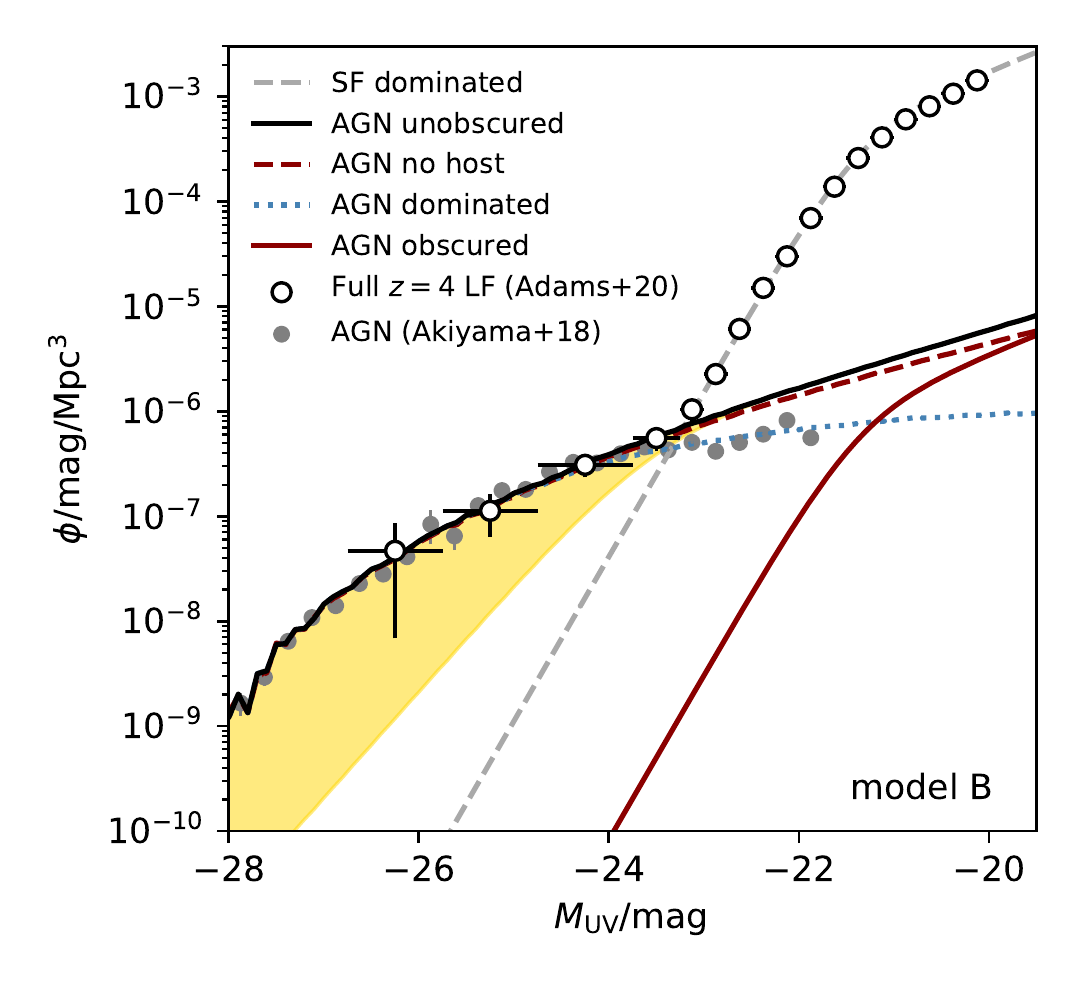}
\caption{
The predicted rest-frame UV LF of AGN as derived from our simple models.
The upper plot shows the results for a model where $m_{\rm BH}/M_{\star} = 0.05$ (model A), and the lower plot shows the results where $m_{\rm BH} \propto M_{\star}^{1.4}$ (model B).
The grey dashed line shows the DPL fit to the rest-frame UV LF for star-forming galaxies from~\citet{Adams2020}.
The result of applying empirical scaling relations, with scatter, to this galaxy LF results in the AGN prediction shown as the black line.
The effect of scatter is highlighted as the gold shaded region, such that the lower edge of this shading would be the prediction with no scatter.
The red solid line shows the predicted obscured Type II AGN LF, and the blue dotted line shows the expected LF for sources in which $L_{\rm BH}/L_{\rm SF} > 12.5$.
The red dashed line shows the predicted AGN LF without the UV emission from the host galaxy.
The open circles show the measurements from~\citet{Adams2020} and the grey filled circles show the results from~\citet{Akiyama2018}, who imposed a criterion on the source morphology.
}
\label{fig:model}
\end{center}
\end{figure}

Following these steps resulted in a simulated AGN LF with a relative flat shape at the bright-end.
This is in contrast with the knee in the function around $M_{\star} \sim -26$ found in observations.
This effect arises in our model from the creation of infeasibly massive black holes and galaxies, due to the application of scatter in the relations between $M_{\rm UV}$, $M_{\star}$ and $m_{\rm BH}$.
While the form of the bright-end of the AGN LF does not impact the results of this work, we nevertheless impose a crude cut in the stellar and black-hole masses to remove these unrealistic sources.
We take a limiting stellar mass of ${\rm log}_{10}(M_{\star}) = 10.8$ from the characteristic mass of high-redshift galaxies (e.g.~\citealp{Ilbert2013, McLeod2020}), and impose an upper limit on the black-hole mass of ${\rm log}_{10}(m_{\rm BH}) = 9.5$ to approximate the drop in the (uncertain) black-hole mass functions at $z \simeq 4$ (e.g.~\citealp{Shankar2009, Kelly2012}).
If a model galaxy/black hole exceeds these mass limits due to scatter, we allocate a lower mass at random according to the relations described above.
The result of this simple process is a knee in the AGN LF in good agreement with the observations.
We calculate the results of this analysis for both the Schechter and DPL form presented in~\citet{Adams2020}, where they fitted to only points at $M_{\rm UV} > -22.0$ to ensure that the results were not influenced by AGN-dominated sources.
The resulting AGN LF depends only weakly on the assumed shape of the LBG LF because the majority of the simulated AGN are hosted in galaxies with $M_{UV} > -22.0$ where there is good agreement between the Schechter and DPL fits.
In contrast, the~\emph{ratio} of AGN to SF-dominated sources at $M_{\rm UV} \sim -23$ does rely on the LBG LF shape, as it depends on how steeply the LBG LF drops-off at the bright-end (see Section~\ref{sect:dis}).

To obtain a predicted AGN LF that matches the number density of quasars known at $z \simeq 4$ we include two factors to modulate the number of LBGs that host an AGN.
The first is the obscured fraction, which describes how many `on' AGN are not bright in the rest-frame UV continuum (e.g. are obscured Type II AGN).
We fixed $f_{\rm obsc.} = 0.6$~\citep{Ueda2014, Vito2018}.
The second modulating factor is the fraction of galaxies that host an active black hole, which is a proxy for the duty cycle of AGN activity.
The $f_{\rm active}$ was determined in our model as the factor required to bring the predicted AGN LF in agreement with the observed AGN LF in the range $-27 < M_{\rm UV} < -24$.
We find $f_{\rm active} = 0.0007$ for model A and $f_{\rm active} = 0.003$ for model B.
Note that $f_{\rm active}$ is not directly comparable to the duty cycle of AGN activity, as it is the fraction of AGN that appear bright in the rest-frame UV rather than the fraction of active black holes.

\subsection{Results}

\begin{figure}
\begin{center}
\includegraphics[width =0.48\textwidth]{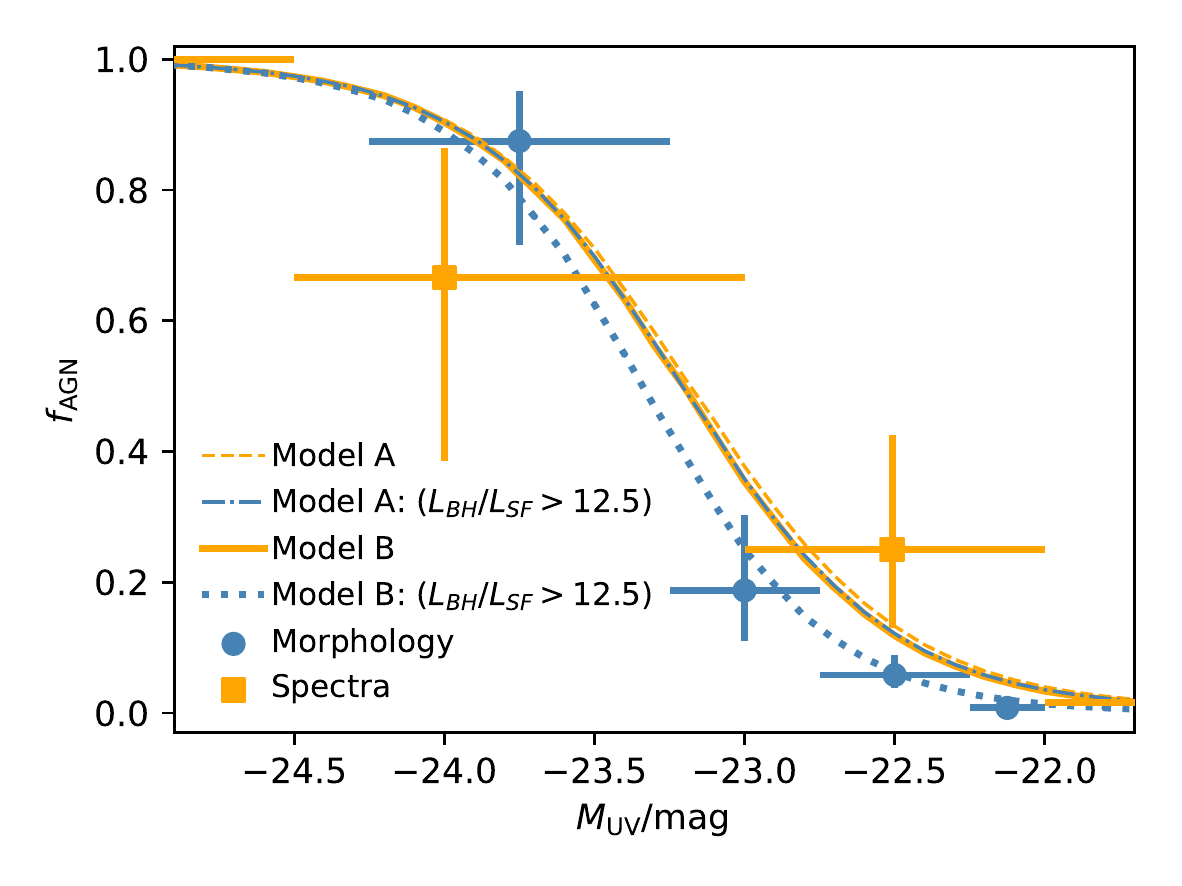}
\caption{
The predicted AGN fraction as a function of absolute UV magnitude at $z = 4$ from our toy model.
The models are compared to the derived fraction from our morphology (blue circles) and spectroscopy (orange squares) data.
The yellow dashed (solid) lines show the predicted $f_{\rm AGN}$ from model A (B) with all AGN included.
The blue dot-dashed (dotted) line corresponds to the $f_{\rm AGN}$ if a luminosity cut of $L_{\rm BH}/L_{\rm SF}$ is imposed to model A (B), to identify AGN-dominated sources.
We see that with this cut model B is able to reproduce our observed morphology-based AGN fraction.
}
\label{fig:agnfracmodel}
\end{center}
\end{figure}

We present the results of this analysis in comparison with the observed LBG and AGN rest-frame UV LF in Fig.~\ref{fig:model}.
Despite the simplicity of the model, it does a reasonable job at reproducing the shape of the observed $ z \simeq 4$ AGN LF at luminosities bright-ward of $M_{\rm UV} \simeq -23.5$.
A striking feature of the predicted LFs shown in Fig.~\ref{fig:model} is the dominant role of scatter which is well known to be important from observations of the brightest quasars (e.g.~\citealp{Venemans2017, Willott2013quasars, Targett2012}).
Furthermore, both models predict a similarly steep faint-end slope consistent with that found by other empirical predictions (e.g.~\citealp{Veale2014,Ren2020, Delvecchio2020}).
For sources around $M_{\rm UV} \simeq -23$, our two models give different predictions for the importance of scatter in the observed galaxies.
This has consequences for the expected morphology and spectroscopic properties of sources around this magnitude.
The AGN LFs from our toy model do not show the flattening observed in the results of~\citet{Akiyama2018} faint-ward of $M_{\rm UV} \simeq -24$.
\citet{Akiyama2018} used a moment based measure to select compact sources in ground-based HSC data, while our model does not make any assumption about the size or morphology of each model galaxy/AGN.
If we impose the condition that the AGN luminosity must be a certain multiple of the stellar light (e.g. $L_{\rm BH}/L_{\rm SF} \gtrsim 10$) then we are able to reproduce the flattening found in the data, but only in the case where the black-hole mass is an increasing fraction of the stellar mass (model B).
The magnitude of the cut is motivated by previous studies comparing the host galaxy and AGN emission in the rest-frame UV and optical for sources selected as quasars (typically $\Delta M = 2$--$3\,{\rm mag}$; ~\citealp{Jahnke2004, Schramm2008, Goto2009, Mechtley2016, Lawther2018})
When fitting the~\citet{Akiyama2018} points with the ratio as a free parameter, we found a best-fit ratio of $L_{\rm BH}/L_{\rm SF} = 12.5 \pm 0.5$ for model B.
The drop in the AGN LF in this case is due to the host galaxy rest-frame UV light becoming significant at $M_{\rm UV} > -23$.
Such a cut is incomplete to fainter AGN (relative to their host galaxy UV emission) and therefore results in an artificial flattening in the observed AGN LF as shown in Fig.~\ref{fig:model}.
If instead black holes populate host galaxies as in our model A, where the effect of scatter in shaping the observed AGN LF at fainter magnitudes is dominant, then such a flattening is not predicted with this relative luminosity cut.
The effect can also be seen in Fig.~\ref{fig:agnfracmodel} where we compare the observed $f_{\rm AGN}$ to that predicted from our model.
These results demonstrate that the morphological and spectroscopic properties of sources around $M_{\rm UV} \simeq -23$ gives important information about how active black holes are distributed within host galaxies.
If the rest-frame UV is always dominated by light from the AGN, then selections based on a point-sources condition will be complete (model A).
Such a selection will not be feasible to fainter magnitudes however, due to LBGs themselves becoming more compact.
If instead the host galaxy light can become important at fainter magnitudes, as in our model B, then we see a distinct incompleteness in point-source selections for AGN at $M_{\rm UV} > -23$.
In this case it becomes essential to define the relative `AGN-strength' that is being included with a given selection methodology, to fully understand what population is being measured.

\subsection{Obscured Type-II AGN}\label{sect:typeII}
So far in this work we have considered only unobscured Type I AGN, which we expect to contribute significantly to the rest-frame UV continuum luminosity of the source.
In the orientation-based unified model of AGN we also expect obscured Type II-like AGN, where the presence of an AGN is only observable in the UV via narrow emission lines.
In our model of the AGN LF as derived from the LBG LF, we can predict the number density of these obscured sources.
Interestingly, our preferred model (model B), which can better explain our observations of the $f_{\rm AGN}$ and the observed flattening of the faint-end slope found by~\citet{Akiyama2018}, also predicts an increased contribution of obscured AGN at $M_{\rm UV} > -22$.
This is a natural consequence of our assumed active and obscured fractions of galaxies in the model.
In model B we expect to see an increased contribution from obscured AGN at fainter magnitudes, with the number densities of `obscured' and `unobscured' sources becoming comparable (within a factor of $< 5$) at $M_{\rm UV} \gtrsim -21$.
While there are many assumptions and uncertainties in this prediction, we note that faint-ward of $M_{\rm UV} = -22$ we detect one source with clear AGN signatures in the compilation of spectra for our sample.
This source, ID520330, appears to be an obscured Type II source (Fig.~\ref{fig:faintspectra}) at $M_{\rm UV} = -21$.
From this one object we estimate the number density of obscured AGN at this magnitude is around $\phi = 7 \pm 7 \times 10^{-6}\,/{\rm mag}/{\rm Mpc}^{3}$, which despite the huge uncertainties is within a factor of ten from our model prediction (model B: Fig.~\ref{fig:model}). 
These arguments demonstrate that there is still considerable uncertainty in the faint-end of the $z \simeq 4$ AGN UV LF depending on how AGN are defined and on the selection procedure.
Given the large samples of $z \simeq 4$ sources available to-date from deep optical/NIR surveys (e.g.~\citealp{Adams2020, Ono2018, Bouwens2015}), the next steps to overcome this challenge do not require substantial increases in sample size.
Rather, in this work we have demonstrated that with a combination of magnitude limited spectroscopic follow-up, coupled with high-resolution imaging, it will be possible to probe the connection between faint-AGN and their galaxy hosts.

\section{Conclusions}\label{sect:conc}
We present the size, morphology and spectroscopic properties of a sample of $3.5 < z <4.5$ galaxies and AGN selected based on a photometric redshift fitting analysis in~\citet{Adams2020}.
The broad magnitude range probed by the parent sample ($-26 \lesssim M_{\rm UV} \lesssim -20$) allows us to uniquely probe the transition between SF- and AGN-dominated sources.
We use both ground-based and~\emph{HST} imaging data to identify the changes in morphology and size, and archival spectra to detect signatures of AGN in the rest-frame UV spectrum.
The key conclusions of this study are as follows.

\begin{itemize}
\item We find the expected galaxy size-luminosity relation up to an absolute UV magnitude of $M_{\rm UV} = -22.5$, beyond which we observe a steep downturn due to the increasing presence of objects with a point-source morphology.  The effect is seen in both the high-resolution~\emph{HST} $I_{814}$ imaging and the ground-based data.
We find that brightest galaxies in the sample have a highly irregular structure as expected from previous works.
\item The existence of archival spectra for a sub-set of our sample allows us to identify SF and AGN dominated sources from the rest-frame UV spectral signatures.  At the bright-end of our sample we see clear AGN signatures in the available spectra, while deep spectroscopy from targeted high-redshift surveys show the expected features of LBGs.  We identify a very bright source $M_{\rm UV} = -23.6$ ($SFR \simeq 80\,{\rm M}_{\odot}/{\rm yr}$) that shows no evidence for an AGN contribution to the rest-frame UV light. 
\item We combine the morphology/size and spectroscopy information to estimate the AGN fraction as a function of $M_{\rm UV}$.  We find a steep transition at $M_{\rm UV} \simeq -23.2$ where the number of bright galaxies drops while AGN-dominated sources become ubiquitous.
We find a slight tension in the $f_{\rm AGN}$ derived independently from our morphology and spectroscopy data at $M_{\rm UV} \simeq -22.5$, with the spectroscopy results finding a higher fraction by a factor of $\sim 5$.
\item We use this AGN fraction to estimate the separated AGN and SF-dominated rest-frame UV LFs at $z \simeq 4$.
We find the bright-end of the SF-dominated LF to be described by a DPL with a bright-end slope of $\beta = -5.15 \pm 0.10$.
Our LBG UV LF is consistent with that expected from the observed steepening in $\beta$ from $z \simeq 9$--$5$ found by~\citet{Bowler2020}, which can be explained by an increased effect of dust attenuation in the most highly star-forming galaxies.
\item We find that the slope of the faint-end of the AGN LF depends on how we determine the AGN fraction.  If we impose a point-source morphology criterion, as in several recent studies of faint AGN, then we find a shallow slope with $\alpha = -1.19 \pm 0.05$.  Conversely, if we derive the AGN number density using the spectroscopic results we find a steeper slope of $\alpha = -1.83 \pm 0.11$.
\item A simple model of the AGN LF, derived using empirical relations applied to the LBG UV LF at $z = 4$, can provide a good description of the transition from AGN to SF-dominated sources.
By applying a criterion on the relative emission from the AGN and host galaxy ($L_{\rm BH}/L_{\rm SF} > 15$), we are able to reproduce the observed flattening of the $ z = 4$ AGN LF at $M_{\rm UV} < -22$ found by~\citet{Akiyama2018}.  
This flattening is only predicted in the case that the light from SF becomes significant in comparison to the AGN in less massive galaxies.

\end{itemize}

Our results demonstrate that while the increasingly large samples of $z \simeq 4$ sources have resulted in low statistical errors on the rest-frame UV LF of AGN, there remain considerable systemic uncertainties on the faint-end of this function.
In particular, the commonly imposed point-source criterion in the selection of AGN samples at these redshifts can result in incomplete samples of active sources at $M_{\rm UV} > -24$ due to the impact of the host galaxy.
The degree of this incompleteness depends on how active black holes populate the underlying galaxy distribution and how these active sources appear in the rest-frame UV light accessible in optical datasets.
Upcoming wide-area high-resolution imaging (e.g. from~\emph{Euclid};~\citealp{Laureijs2012}) with extensive spectroscopic follow-up (e.g. from degree-scale multi-object spectrographs like the William Herschel Telescope Enhanced Area Velocity Explorer;~\citealp{Dalton2012} and the Multi-Object Optical and Near-infrared Spectrograph;~\citealp{Cirasuolo2014}) will be a powerful combination to understand further the co-evolution of galaxies and AGN at high redshifts.

\section*{Acknowledgements}
We acknowledge useful discussions with Fergus Cullen, Paul Hewett, Manda Banerji and the `Quasar Souls' group at the Institute of Astronomy at the University of Cambridge.
We acknowledge Kate Gould for compiling the archival spectra.
We thank the anonymous referee for comments that improved this paper.
This work was supported by the Glasstone Foundation and the Oxford Hintze Centre for Astrophysical Surveys which is funded through generous support from the Hintze Family Charitable Foundation.
Funding for the SDSS and SDSS-II has been provided by the Alfred P. Sloan Foundation, the Participating Institutions, the National Science Foundation, the U.S. Department of Energy, the National Aeronautics and Space Administration, the Japanese Monbukagakusho, the Max Planck Society, and the Higher Education Funding Council for England. The SDSS Web Site is http://www.sdss.org/.
The SDSS is managed by the Astrophysical Research Consortium for the Participating Institutions. The Participating Institutions are the American Museum of Natural History, Astrophysical Institute Potsdam, University of Basel, University of Cambridge, Case Western Reserve University, University of Chicago, Drexel University, Fermilab, the Institute for Advanced Study, the Japan Participation Group, Johns Hopkins University, the Joint Institute for Nuclear Astrophysics, the Kavli Institute for Particle Astrophysics and Cosmology, the Korean Scientist Group, the Chinese Academy of Sciences (LAMOST), Los Alamos National Laboratory, the Max-Planck-Institute for Astronomy (MPIA), the Max-Planck-Institute for Astrophysics (MPA), New Mexico State University, Ohio State University, University of Pittsburgh, University of Portsmouth, Princeton University, the United States Naval Observatory, and the University of Washington. 
 This research uses data from the VIMOS VLT Deep Survey, obtained from the VVDS database operated by Cesam, Laboratoire d'Astrophysique de Marseille, France. 
This research has made use of the zCosmos database, operated at CeSAM/LAM, Marseille, France. 

\section*{Data Availability}

The datasets used in this work were derived from sources in the public domain.
Links to the online repositories and references to the survey data we utilized are listed in Section~\ref{sect:data}.




\bibliographystyle{mnras}
\bibliography{library_abbrv} 

\begin{thebibliography}{}
\makeatletter
\relax
\def\mn@urlcharsother{\let\do\@makeother \do\$\do\&\do\#\do\^\do\_\do\%\do\~}
\def\mn@doi{\begingroup\mn@urlcharsother \@ifnextchar [ {\mn@doi@}
  {\mn@doi@[]}}
\def\mn@doi@[#1]#2{\def\@tempa{#1}\ifx\@tempa\@empty \href
  {http://dx.doi.org/#2} {doi:#2}\else \href {http://dx.doi.org/#2} {#1}\fi
  \endgroup}
\def\mn@eprint#1#2{\mn@eprint@#1:#2::\@nil}
\def\mn@eprint@arXiv#1{\href {http://arxiv.org/abs/#1} {{\tt arXiv:#1}}}
\def\mn@eprint@dblp#1{\href {http://dblp.uni-trier.de/rec/bibtex/#1.xml}
  {dblp:#1}}
\def\mn@eprint@#1:#2:#3:#4\@nil{\def\@tempa {#1}\def\@tempb {#2}\def\@tempc
  {#3}\ifx \@tempc \@empty \let \@tempc \@tempb \let \@tempb \@tempa \fi \ifx
  \@tempb \@empty \def\@tempb {arXiv}\fi \@ifundefined
  {mn@eprint@\@tempb}{\@tempb:\@tempc}{\expandafter \expandafter \csname
  mn@eprint@\@tempb\endcsname \expandafter{\@tempc}}}

\bibitem[\protect\citeauthoryear{Adams, Bowler, Jarvis, H{\"{a}}u{\ss}ler,
  McLure, Bunker, Dunlop  \& Verma}{Adams et~al.}{2020}]{Adams2020}
Adams N.~J.,  Bowler R. A.~A.,  Jarvis M.~J.,  H{\"{a}}u{\ss}ler B.,  McLure
  R.~J.,  Bunker A.,  Dunlop J.~S.,   Verma A.,  2020, \mn@doi [MNRAS]
  {10.1093/mnras/staa687}, 494, 1771

\bibitem[\protect\citeauthoryear{Akiyama et~al.,}{Akiyama
  et~al.}{2018}]{Akiyama2018}
Akiyama M.,  et~al., 2018, \mn@doi [PASJ] {10.1093/pasj/psx091}, 70, S34

\bibitem[\protect\citeauthoryear{Alexandroff et~al.,}{Alexandroff
  et~al.}{2013}]{Alexandroff2013}
Alexandroff R.,  et~al., 2013, \mn@doi [MNRAS] {10.1093/mnras/stt1500}, 435,
  3306

\bibitem[\protect\citeauthoryear{Ba{\~{n}}ados et~al.,}{Ba{\~{n}}ados
  et~al.}{2016}]{Banados2016}
Ba{\~{n}}ados E.,  et~al., 2016, \mn@doi [ApJSS] {10.3847/0067-0049/227/1/11},
  227, 11

\bibitem[\protect\citeauthoryear{Ba{\~{n}}ados et~al.,}{Ba{\~{n}}ados
  et~al.}{2018}]{Banados2018}
Ba{\~{n}}ados E.,  et~al., 2018, \mn@doi [Nat] {10.1038/nature25180}, 553, 473

\bibitem[\protect\citeauthoryear{Barone-Nugent, Wyithe, Trenti, Treu, Oesch,
  Bouwens, Illingworth  \& Schmidt}{Barone-Nugent
  et~al.}{2015}]{Barone-Nugent2015a}
Barone-Nugent R.~L.,  Wyithe J. S.~B.,  Trenti M.,  Treu T.,  Oesch P.,
  Bouwens R.,  Illingworth G.~D.,   Schmidt K.~B.,  2015, \mn@doi [MNRAS]
  {10.1093/mnras/stv633}, 450, 1224

\bibitem[\protect\citeauthoryear{Bertin \& Arnouts}{Bertin \&
  Arnouts}{1996}]{Bertin1996}
Bertin E.,  Arnouts S.,  1996, A{\&}AS, 117, 393

\bibitem[\protect\citeauthoryear{Bian et~al.,}{Bian et~al.}{2013}]{Bian2013}
Bian F.,  et~al., 2013, \mn@doi [ApJ] {10.1088/0004-637X/774/1/28}, 774, 28

\bibitem[\protect\citeauthoryear{Boutsia, Grazian, Giallongo, Fiore  \&
  Civano}{Boutsia et~al.}{2018}]{Boutsia2018}
Boutsia K.,  Grazian A.,  Giallongo E.,  Fiore F.,   Civano F.,  2018, \mn@doi
  [ApJ] {10.3847/1538-4357/aae6c7}, 869, 20

\bibitem[\protect\citeauthoryear{Bouwens et~al.,}{Bouwens
  et~al.}{2014}]{Bouwens2014beta}
Bouwens R.~J.,  et~al., 2014, \mn@doi [ApJ] {10.1088/0004-637X/793/2/115}, 793,
  115

\bibitem[\protect\citeauthoryear{Bouwens et~al.,}{Bouwens
  et~al.}{2015}]{Bouwens2015}
Bouwens R.~J.,  et~al., 2015, \mn@doi [ApJ] {10.1088/0004-637X/803/1/34}, 803,
  34

\bibitem[\protect\citeauthoryear{Bower, Benson  \& Crain}{Bower
  et~al.}{2012}]{Bower2012}
Bower R.~G.,  Benson A.~J.,   Crain R.~A.,  2012, \mn@doi [MNRAS]
  {10.1111/j.1365-2966.2012.20516.x}, 422, 2816

\bibitem[\protect\citeauthoryear{Bowler et~al.,}{Bowler
  et~al.}{2012}]{Bowler2012}
Bowler R. A.~A.,  et~al., 2012, \mn@doi [MNRAS]
  {10.1111/j.1365-2966.2012.21904.x}, 426, 2772

\bibitem[\protect\citeauthoryear{Bowler et~al.,}{Bowler
  et~al.}{2014}]{Bowler2014}
Bowler R. A.~A.,  et~al., 2014, \mn@doi [MNRAS] {10.1093/mnras/stu449}, 440,
  2810

\bibitem[\protect\citeauthoryear{Bowler et~al.,}{Bowler
  et~al.}{2015}]{Bowler2015}
Bowler R. A.~A.,  et~al., 2015, \mn@doi [MNRAS] {10.1093/mnras/stv1403}, 452,
  1817

\bibitem[\protect\citeauthoryear{Bowler, Dunlop, McLure  \& McLeod}{Bowler
  et~al.}{2017}]{Bowler2017}
Bowler R.,  Dunlop J.,  McLure R.,   McLeod D.,  2017, \mn@doi [MNRAS]
  {10.1093/mnras/stw3296}, 466, 3612

\bibitem[\protect\citeauthoryear{Bowler, Jarvis, Dunlop, McLure, McLeod, Adams,
  Milvang-Jensen  \& McCracken}{Bowler et~al.}{2020}]{Bowler2020}
Bowler R. A.~A.,  Jarvis M.~J.,  Dunlop J.~S.,  McLure R.~J.,  McLeod D.~J.,
  Adams N.~J.,  Milvang-Jensen B.,   McCracken H.~J.,  2020, \mn@doi [MNRAS]
  {10.1093/mnras/staa313}, 2084, 2059

\bibitem[\protect\citeauthoryear{Cirasuolo}{Cirasuolo}{2014}]{Cirasuolo2014}
Cirasuolo M.,  2014, SPIE, 9147, 91470N

\bibitem[\protect\citeauthoryear{Clay, Thomas, Wilkins  \& Henriques}{Clay
  et~al.}{2015}]{Clay2015}
Clay S.,  Thomas P.,  Wilkins S.,   Henriques B.,  2015, MNRAS, 415, 2692

\bibitem[\protect\citeauthoryear{Coil et~al.,}{Coil et~al.}{2011}]{Coil2011}
Coil A.~L.,  et~al., 2011, \mn@doi [ApJ] {10.1088/0004-637X/741/1/8}, 741, 8

\bibitem[\protect\citeauthoryear{Cowie, Barger, Bauer  \&
  Gonz{\'{a}}lez-L{\'{o}}pez}{Cowie et~al.}{2020}]{Cowie2020}
Cowie L.~L.,  Barger A.~J.,  Bauer F.~E.,   Gonz{\'{a}}lez-L{\'{o}}pez J.,
  2020, \mn@doi [ApJ] {10.3847/1538-4357/ab6aaa}, 891, 69

\bibitem[\protect\citeauthoryear{Croton}{Croton}{2006}]{Croton2006}
Croton D.~J.,  2006, \mn@doi [MNRAS] {10.1111/j.1365-2966.2006.10429.x}, 369,
  1808

\bibitem[\protect\citeauthoryear{Curtis-Lake et~al.,}{Curtis-Lake
  et~al.}{2016}]{Curtis-Lake2016}
Curtis-Lake E.,  et~al., 2016, \mn@doi [MNRAS] {10.1093/mnras/stv3017}, 457,
  440

\bibitem[\protect\citeauthoryear{Curtis-Lake, Chevallard  \&
  Charlot}{Curtis-Lake et~al.}{2020}]{CurtisLake2020}
Curtis-Lake E.,  Chevallard J.,   Charlot S.,  2020, preprint
  (arXiv:2001.08560)

\bibitem[\protect\citeauthoryear{Dalton et~al.,}{Dalton
  et~al.}{2012}]{Dalton2012}
Dalton G.,  et~al., 2012, \mn@doi [SPIE] {10.1117/12.925950}, 8446, 84460P

\bibitem[\protect\citeauthoryear{Dayal, Ferrara, Dunlop  \& Pacucci}{Dayal
  et~al.}{2014}]{Dayal2014}
Dayal P.,  Ferrara A.,  Dunlop J.~S.,   Pacucci F.,  2014, \mn@doi [MNRAS]
  {10.1093/mnras/stu1848}, 445, 2545

\bibitem[\protect\citeauthoryear{Delvecchio et~al.,}{Delvecchio
  et~al.}{2020}]{Delvecchio2020}
Delvecchio I.,  et~al., 2020, \mn@doi [ApJ] {10.3847/1538-4357/ab789c}, 892, 17

\bibitem[\protect\citeauthoryear{Duncan et~al.,}{Duncan
  et~al.}{2014}]{Duncan2014}
Duncan K.,  et~al., 2014, \mn@doi [MNRAS] {10.1093/mnras/stu1622}, 444, 2960

\bibitem[\protect\citeauthoryear{Eisenstein et~al.,}{Eisenstein
  et~al.}{2011}]{Eisenstein2011}
Eisenstein D.~J.,  et~al., 2011, \mn@doi [AJ] {10.1088/0004-6256/142/3/72},
  142, 24

\bibitem[\protect\citeauthoryear{Fan et~al.,}{Fan et~al.}{2003}]{Fan2003}
Fan X.,  et~al., 2003, \mn@doi [AJ] {10.1086/368246}, 125, 1649

\bibitem[\protect\citeauthoryear{F{\`{e}}vre, B{\'{e}}thermin, Faisst, Capak,
  Cassata, Silverman, Schaerer  \& Yan}{F{\`{e}}vre et~al.}{2019}]{Fevre2019}
F{\`{e}}vre O.~L.,  B{\'{e}}thermin M.,  Faisst A.,  Capak P.,  Cassata P.,
  Silverman J.~D.,  Schaerer D.,   Yan L.,  2019, preprint(arXiv:1910.09517)

\bibitem[\protect\citeauthoryear{Finkelstein et~al.,}{Finkelstein
  et~al.}{2015}]{Finkelstein2015}
Finkelstein S.~L.,  et~al., 2015, \mn@doi [ApJ] {10.1088/0004-637X/810/1/71},
  810, 71

\bibitem[\protect\citeauthoryear{Gavignaud et~al.,}{Gavignaud
  et~al.}{2006}]{Gavignaud2006}
Gavignaud I.,  et~al., 2006, \mn@doi [A\&A] {10.1051/0004-6361:20065376}, 457,
  79

\bibitem[\protect\citeauthoryear{Giallongo et~al.,}{Giallongo
  et~al.}{2015}]{Giallongo2015}
Giallongo E.,  et~al., 2015, \mn@doi [A\&A] {10.1051/0004-6361/201425334}, 578,
  A83

\bibitem[\protect\citeauthoryear{Giallongo et~al.,}{Giallongo
  et~al.}{2019}]{Giallongo2019}
Giallongo E.,  et~al., 2019, \mn@doi [ApJ] {10.3847/1538-4357/ab39e1}, 884, 19

\bibitem[\protect\citeauthoryear{Gonzalez-Perez, Lacey, Baugh, Frenk  \&
  Wilkins}{Gonzalez-Perez et~al.}{2013}]{GonzalezPerez2013}
Gonzalez-Perez V.,  Lacey C.~G.,  Baugh C.~M.,  Frenk C.~S.,   Wilkins S.~M.,
  2013, \mn@doi [MNRAS] {10.1093/mnras/sts446}, 429, 1609

\bibitem[\protect\citeauthoryear{Goto, Utsumi, Furusawa, Miyazaki  \&
  Komiyama}{Goto et~al.}{2009}]{Goto2009}
Goto T.,  Utsumi Y.,  Furusawa H.,  Miyazaki S.,   Komiyama Y.,  2009, \mn@doi
  [MNRAS] {10.1111/j.1365-2966.2009.15486.x}, 400, 843

\bibitem[\protect\citeauthoryear{Grogin et~al.,}{Grogin
  et~al.}{2011}]{Grogin2011}
Grogin N.~A.,  et~al., 2011, \mn@doi [ApJS] {10.1088/0067-0049/197/2/35}, 197,
  35

\bibitem[\protect\citeauthoryear{Hasinger et~al.,}{Hasinger
  et~al.}{2018}]{Hasinger2018}
Hasinger G.,  et~al., 2018, \mn@doi [ApJ] {10.3847/1538-4357/aabacf}, 858, 77

\bibitem[\protect\citeauthoryear{Hathi et~al.,}{Hathi et~al.}{2010}]{Hathi2010}
Hathi N.~P.,  et~al., 2010, \mn@doi [ApJ] {10.1088/0004-637X/720/2/1708}, 720,
  1708

\bibitem[\protect\citeauthoryear{Hirschmann, Khochfar, Burkert, Naab, Genel  \&
  Somerville}{Hirschmann et~al.}{2010}]{Hirschmann2010}
Hirschmann M.,  Khochfar S.,  Burkert A.,  Naab T.,  Genel S.,   Somerville
  R.~S.,  2010, \mn@doi [MNRAS] {10.1111/j.1365-2966.2010.17006.x}, 407, 1016

\bibitem[\protect\citeauthoryear{Huang, Ferguson, Ravindranath  \& Su}{Huang
  et~al.}{2013}]{Huang2013}
Huang K.-H.,  Ferguson H.~C.,  Ravindranath S.,   Su J.,  2013, \mn@doi [ApJ]
  {10.1088/0004-637X/765/1/68}, 765, 68

\bibitem[\protect\citeauthoryear{Ikeda et~al.,}{Ikeda et~al.}{2012}]{Ikeda2012}
Ikeda H.,  et~al., 2012, \mn@doi [ApJ] {10.1088/0004-637X/756/2/160}, 756, 160

\bibitem[\protect\citeauthoryear{Ilbert et~al.,}{Ilbert
  et~al.}{2013}]{Ilbert2013}
Ilbert O.,  et~al., 2013, \mn@doi [A{\&}A] {10.1051/0004-6361/201321100}, 556,
  A55

\bibitem[\protect\citeauthoryear{Jahnke et~al.,}{Jahnke
  et~al.}{2004}]{Jahnke2004}
Jahnke K.,  et~al., 2004, \mn@doi [ApJ] {10.1086/423233}, 614, 568

\bibitem[\protect\citeauthoryear{Jiang et~al.,}{Jiang
  et~al.}{2013}]{Jiang2013b}
Jiang L.,  et~al., 2013, \mn@doi [ApJ] {10.1088/0004-637X/773/2/153}, 773, 153

\bibitem[\protect\citeauthoryear{Jiang et~al.,}{Jiang et~al.}{2016}]{Jiang2016}
Jiang L.,  et~al., 2016, \mn@doi [ApJ] {10.3847/1538-4357/833/2/222}, 833

\bibitem[\protect\citeauthoryear{Kashikawa et~al.,}{Kashikawa
  et~al.}{2015}]{Kashikawa2014a}
Kashikawa N.,  et~al., 2015, \mn@doi [ApJ] {10.1088/0004-637X/798/1/28}, 798,
  28

\bibitem[\protect\citeauthoryear{Kelly \& Merloni}{Kelly \&
  Merloni}{2012}]{Kelly2012}
Kelly B.~C.,  Merloni A.,  2012, \mn@doi [Adv. in Ast.] {10.1155/2012/970858},
  2012

\bibitem[\protect\citeauthoryear{Kelly \& Shen}{Kelly \&
  Shen}{2013}]{Kelly2013}
Kelly B.~C.,  Shen Y.,  2013, \mn@doi [ApJ] {10.1088/0004-637X/764/1/45}, 764

\bibitem[\protect\citeauthoryear{Kim et~al.,}{Kim et~al.}{2019}]{Kim2019}
Kim Y.,  et~al., 2019, \mn@doi [ApJ] {10.3847/1538-4357/aaf387}, 870, 86

\bibitem[\protect\citeauthoryear{Koekemoer et~al.,}{Koekemoer
  et~al.}{2007}]{Koekemoer2007}
Koekemoer A.~M.,  et~al., 2007, \mn@doi [ApJS] {10.1086/520086}, 172, 196

\bibitem[\protect\citeauthoryear{Koekemoer et~al.,}{Koekemoer
  et~al.}{2011}]{Koekemoer2011}
Koekemoer A.~M.,  et~al., 2011, \mn@doi [ApJS] {10.1088/0067-0049/197/2/36},
  197, 36

\bibitem[\protect\citeauthoryear{Laporte et~al.,}{Laporte
  et~al.}{2017}]{Laporte2017}
Laporte N.,  et~al., 2017, \mn@doi [ApJ] {10.3847/2041-8213/aa62aa}, 837, L21

\bibitem[\protect\citeauthoryear{Laureijs et~al.,}{Laureijs
  et~al.}{2012}]{Laureijs2012}
Laureijs R.,  et~al., 2012, in Clampin M.~C.,  Fazio G.~G.,  MacEwen H.~A.,
  Oschmann J.~M.,  eds,  Vol. 8442, Space Telescopes and Instrumentation 2012:
  Optical. p. 84420T, \mn@doi{10.1117/12.926496}, \url
  {http://adsabs.harvard.edu/abs/2012SPIE.8442E..0TL}

\bibitem[\protect\citeauthoryear{Law, Steidel, Shapley, Nagy, Reddy  \&
  Erb}{Law et~al.}{2012}]{Law2012}
Law D.~R.,  Steidel C.~C.,  Shapley A.~E.,  Nagy S.~R.,  Reddy N.~A.,   Erb
  D.~K.,  2012, \mn@doi [ApJ] {10.1088/0004-637X/745/1/85}, 745, 85

\bibitem[\protect\citeauthoryear{Lawther, Vestergaard  \& Fan}{Lawther
  et~al.}{2018}]{Lawther2018}
Lawther D.,  Vestergaard M.,   Fan X.,  2018, \mn@doi [MNRAS]
  {10.1093/mnras/stx3203}, 475, 3213

\bibitem[\protect\citeauthoryear{{Le F{\`{e}}vre} et~al.,}{{Le F{\`{e}}vre}
  et~al.}{2015}]{LeFevre2015}
{Le F{\`{e}}vre} O.,  et~al., 2015, \mn@doi [A\&A]
  {10.1051/0004-6361/201423829}, 576, A79

\bibitem[\protect\citeauthoryear{Lee et~al.,}{Lee et~al.}{2011}]{Lee2011}
Lee K.-s.,  et~al., 2011, \mn@doi [ApJ] {10.1088/0004-637X/733/2/99}, 733, 99

\bibitem[\protect\citeauthoryear{Lilly et~al.,}{Lilly et~al.}{2007}]{Lilly2007}
Lilly S.~J.,  et~al., 2007, \mn@doi [ApJS] {10.1086/516589}, 172, 70

\bibitem[\protect\citeauthoryear{Lotz, Madau, Giavalisco, Primack  \&
  Ferguson}{Lotz et~al.}{2006}]{Lotz2006}
Lotz J.~M.,  Madau P.,  Giavalisco M.,  Primack J.,   Ferguson H.~C.,  2006,
  \mn@doi [ApJ] {10.1086/497950}, 636, 592

\bibitem[\protect\citeauthoryear{Madau, Pozzetti  \& Dickinson}{Madau
  et~al.}{1998}]{Madau1998}
Madau P.,  Pozzetti L.,   Dickinson M.,  1998, \mn@doi [ApJ] {10.1086/305523},
  498, 106

\bibitem[\protect\citeauthoryear{Mason et~al.,}{Mason et~al.}{2015}]{Mason2015}
Mason C.~A.,  et~al., 2015, \mn@doi [ApJ] {10.1088/0004-637X/805/1/79}, 805, 79

\bibitem[\protect\citeauthoryear{Massey, Stoughton, Leauthaud, Rhodes,
  Koekemoer, Ellis  \& Shaghoulian}{Massey et~al.}{2010}]{Massey2010}
Massey R.,  Stoughton C.,  Leauthaud A.,  Rhodes J.,  Koekemoer A.,  Ellis R.,
   Shaghoulian E.,  2010, \mn@doi [MNRAS] {10.1111/j.1365-2966.2009.15638.x},
  401, 371

\bibitem[\protect\citeauthoryear{Masters et~al.,}{Masters
  et~al.}{2012}]{Masters2012}
Masters D.,  et~al., 2012, \mn@doi [ApJ] {10.1088/2041-8205/752/1/L14}, 752,
  L14

\bibitem[\protect\citeauthoryear{Matsuoka et~al.,}{Matsuoka
  et~al.}{2018a}]{Matsuoka2018}
Matsuoka Y.,  et~al., 2018a, \mn@doi [ApJS] {10.3847/1538-4365/aac724}, 237, 5

\bibitem[\protect\citeauthoryear{Matsuoka et~al.,}{Matsuoka
  et~al.}{2018b}]{Matsuoka2018a}
Matsuoka Y.,  et~al., 2018b, \mn@doi [ApJ] {10.3847/1538-4357/aaee7a}, 869, 150

\bibitem[\protect\citeauthoryear{Matute, Masegosa, M{\'{a}}rquez, Husillos,
  Olmo, Perea  \& Povi}{Matute et~al.}{2013}]{Matute2013}
Matute I.,  Masegosa J.,  M{\'{a}}rquez I.,  Husillos C.,  Olmo A.,  Perea J.,
   Povi M.,  2013, A{\&}A, 557, A78

\bibitem[\protect\citeauthoryear{McGreer et~al.,}{McGreer
  et~al.}{2013}]{McGreer2013}
McGreer I.~D.,  et~al., 2013, \mn@doi [ApJ] {10.1088/0004-637X/768/2/105}, 768,
  105

\bibitem[\protect\citeauthoryear{McGreer, Fan, Jiang  \& Cai}{McGreer
  et~al.}{2018}]{McGreer2018}
McGreer I.,  Fan X.,  Jiang L.,   Cai Z.,  2018, \mn@doi [ApJ]
  {10.3847/1538-3881/aaaab4}, 155, 131

\bibitem[\protect\citeauthoryear{McLeod, McLure, Dunlop, Cullen, Carnall  \&
  Duncan}{McLeod et~al.}{2020}]{McLeod2020}
McLeod D.~J.,  McLure R.~J.,  Dunlop J.~S.,  Cullen F.,  Carnall A.~C.,
  Duncan K.,  2020, 24, 1

\bibitem[\protect\citeauthoryear{McLure et~al.,}{McLure
  et~al.}{2013}]{McLure2013}
McLure R.~J.,  et~al., 2013, \mn@doi [MNRAS] {10.1093/mnras/stt627}, 432, 2696

\bibitem[\protect\citeauthoryear{McLure et~al.,}{McLure
  et~al.}{2018}]{McLure2018}
McLure R.~J.,  et~al., 2018, \mn@doi [MNRAS] {10.1093/mnras/sty1213}, 479, 25

\bibitem[\protect\citeauthoryear{Mechtley et~al.,}{Mechtley
  et~al.}{2016}]{Mechtley2016}
Mechtley M.,  et~al., 2016, \mn@doi [ApJ] {10.3847/0004-637x/830/2/156}, 830,
  156

\bibitem[\protect\citeauthoryear{Mortlock et~al.,}{Mortlock
  et~al.}{2011}]{Mortlock2011}
Mortlock D.~J.,  et~al., 2011, \mn@doi [Nat] {10.1038/nature10159}, 474, 616

\bibitem[\protect\citeauthoryear{Oke}{Oke}{1974}]{Oke1974}
Oke J.~B.,  1974, \mn@doi [ApJS] {10.1086/190287}, 27, 21

\bibitem[\protect\citeauthoryear{Oke \& Gunn}{Oke \& Gunn}{1983}]{Oke1983}
Oke J.~B.,  Gunn J.~E.,  1983, \mn@doi [ApJ] {10.1086/160817}, 266, 713

\bibitem[\protect\citeauthoryear{Ono et~al.,}{Ono et~al.}{2018}]{Ono2018}
Ono Y.,  et~al., 2018, \mn@doi [PASJ] {10.1093/pasj/psx103}, 70, S10

\bibitem[\protect\citeauthoryear{Parsa, Dunlop, McLure  \& Mortlock}{Parsa
  et~al.}{2016}]{Parsa2016}
Parsa S.,  Dunlop J.~S.,  McLure R.~J.,   Mortlock A.,  2016, \mn@doi [MNRAS]
  {10.1093/mnras/stv2857}, 456, 3194

\bibitem[\protect\citeauthoryear{Parsa, Dunlop  \& McLure}{Parsa
  et~al.}{2018}]{Parsa2018}
Parsa S.,  Dunlop J.~S.,   McLure R.~J.,  2018, \mn@doi [MNRAS]
  {10.1093/mnras/stx2887}, 474, 2904

\bibitem[\protect\citeauthoryear{Pentericci et~al.,}{Pentericci
  et~al.}{2018}]{Pentericci2018a}
Pentericci L.,  et~al., 2018, \mn@doi [A{\&}A] {10.1051/0004-6361/201833047},
  616, A174

\bibitem[\protect\citeauthoryear{Reines \& Volonteri}{Reines \&
  Volonteri}{2015}]{Reines2015}
Reines A.~E.,  Volonteri M.,  2015, \mn@doi [ApJ] {10.1088/0004-637X/813/2/82},
  813, 82

\bibitem[\protect\citeauthoryear{Ren, Trenti  \& {Di Matteo}}{Ren
  et~al.}{2020}]{Ren2020}
Ren K.,  Trenti M.,   {Di Matteo} T.,  2020, \mn@doi [ApJ]
  {10.3847/1538-4357/ab86ab}, 894, 124

\bibitem[\protect\citeauthoryear{Richards et~al.,}{Richards
  et~al.}{2002}]{Richards2002}
Richards G.~T.,  et~al., 2002, \mn@doi [AJ] {10.1086/340187}, 123, 2945

\bibitem[\protect\citeauthoryear{Richards et~al.,}{Richards
  et~al.}{2006}]{Richards2006}
Richards G.~T.,  et~al., 2006, \mn@doi [AJ] {10.1086/503559}, 131, 2766

\bibitem[\protect\citeauthoryear{Rogers et~al.,}{Rogers
  et~al.}{2014}]{Rogers2014}
Rogers A.~B.,  et~al., 2014, \mn@doi [MNRAS] {10.1093/mnras/stu558}, 440, 3714

\bibitem[\protect\citeauthoryear{Runnoe, Brotherton  \& Shang}{Runnoe
  et~al.}{2012}]{Runnoe2012}
Runnoe J.~C.,  Brotherton M.~S.,   Shang Z.,  2012, \mn@doi [MNRAS]
  {10.1111/j.1365-2966.2012.20620.x}, 422, 478

\bibitem[\protect\citeauthoryear{Salmon et~al.,}{Salmon
  et~al.}{2015}]{Salmon2015}
Salmon B.,  et~al., 2015, \mn@doi [ApJ] {10.1088/0004-637X/799/2/183}, 799, 183

\bibitem[\protect\citeauthoryear{Schramm, Wisotzki  \& Jahnke}{Schramm
  et~al.}{2008}]{Schramm2008}
Schramm M.,  Wisotzki L.,   Jahnke K.,  2008, \mn@doi [A\&A]
  {10.1051/0004-6361:20077319}, 478, 311

\bibitem[\protect\citeauthoryear{Scoville et~al.,}{Scoville
  et~al.}{2007}]{Scoville2007}
Scoville N.,  et~al., 2007, \mn@doi [ApJS] {10.1086/516585}, 172, 1

\bibitem[\protect\citeauthoryear{Shankar, Weinberg  \&
  Miralda-Escud{\'{e}}}{Shankar et~al.}{2009}]{Shankar2009}
Shankar F.,  Weinberg D.~H.,   Miralda-Escud{\'{e}} J.,  2009, \mn@doi [ApJ]
  {10.1088/0004-637X/690/1/20}, 690, 20

\bibitem[\protect\citeauthoryear{Shapley, Steidel, Pettini  \&
  Adelberger}{Shapley et~al.}{2003}]{Shapley2003}
Shapley A.~E.,  Steidel C.~C.,  Pettini M.,   Adelberger K.~L.,  2003, \mn@doi
  [ApJ] {10.1086/373922}, 588, 65

\bibitem[\protect\citeauthoryear{Shen, Mo, White, Blanton, Kauffmann, Voges,
  Brinkmann  \& Csabai}{Shen et~al.}{2003}]{Shen2003}
Shen S.,  Mo H.~J.,  White S. D.~M.,  Blanton M.~R.,  Kauffmann G.,  Voges W.,
  Brinkmann J.,   Csabai I.,  2003, \mn@doi [MNRAS]
  {10.1046/j.1365-8711.2003.06740.x}, 343, 978

\bibitem[\protect\citeauthoryear{Shin et~al.,}{Shin et~al.}{2020}]{Shin2020}
Shin S.,  et~al., 2020, \mn@doi [ApJ] {10.3847/1538-4357/ab7bde}, 893, 45

\bibitem[\protect\citeauthoryear{Sobral et~al.,}{Sobral
  et~al.}{2018}]{Sobral2018}
Sobral D.,  et~al., 2018, \mn@doi [MNRAS] {10.1093/mnras/sty2779}, 482, 2422

\bibitem[\protect\citeauthoryear{Song et~al.,}{Song et~al.}{2016}]{Song2016}
Song M.,  et~al., 2016, \mn@doi [ApJ] {10.3847/0004-637x/825/1/5}, 825, 5

\bibitem[\protect\citeauthoryear{Stark, Schenker, Ellis, Robertson, McLure  \&
  Dunlop}{Stark et~al.}{2013}]{Stark2013}
Stark D.~P.,  Schenker M.~A.,  Ellis R.,  Robertson B.,  McLure R.,   Dunlop
  J.,  2013, \mn@doi [ApJ] {10.1088/0004-637X/763/2/129}, 763, 129

\bibitem[\protect\citeauthoryear{Stark et~al.,}{Stark et~al.}{2014}]{Stark2014}
Stark D.~P.,  et~al., 2014, \mn@doi [MNRAS] {10.1093/mnras/stu1618}, 445, 3200

\bibitem[\protect\citeauthoryear{Steidel, Strom, Pettini, Rudie, Reddy  \&
  Trainor}{Steidel et~al.}{2016}]{Steidel2016}
Steidel C.~C.,  Strom A.~L.,  Pettini M.,  Rudie G.~C.,  Reddy N.~A.,   Trainor
  R.~F.,  2016, \mn@doi [ApJ] {10.3847/0004-637X/826/2/159}, 826, 159

\bibitem[\protect\citeauthoryear{Stevans et~al.,}{Stevans
  et~al.}{2018}]{Stevans2018}
Stevans M.~L.,  et~al., 2018, \mn@doi [ApJ] {10.3847/1538-4357/aacbd7}, 863, 63

\bibitem[\protect\citeauthoryear{Tacchella, Bose, Conroy, Eisenstein  \&
  Johnson}{Tacchella et~al.}{2018}]{Tacchella2018}
Tacchella S.,  Bose S.,  Conroy C.,  Eisenstein D.~J.,   Johnson B.~D.,  2018,
  \mn@doi [ApJ] {10.3847/1538-4357/aae8e0}, 868, 92

\bibitem[\protect\citeauthoryear{Targett, Dunlop  \& Mclure}{Targett
  et~al.}{2012}]{Targett2012}
Targett T.~A.,  Dunlop J.~S.,   Mclure R.~J.,  2012, \mn@doi [MNRAS]
  {10.1111/j.1365-2966.2011.20286.x}, 420, 3621

\bibitem[\protect\citeauthoryear{Tilvi et~al.,}{Tilvi et~al.}{2016}]{Tilvi2016}
Tilvi V.,  et~al., 2016, ApJL, 827, L14

\bibitem[\protect\citeauthoryear{Ueda, Akiyama, Hasinger, Miyaji  \&
  Watson}{Ueda et~al.}{2014}]{Ueda2014}
Ueda Y.,  Akiyama M.,  Hasinger G.,  Miyaji T.,   Watson M.~G.,  2014, \mn@doi
  [ApJ] {10.1088/0004-637X/786/2/104}, 786, 104

\bibitem[\protect\citeauthoryear{{Vanden Berk} et~al.,}{{Vanden Berk}
  et~al.}{2001}]{VandenBerk2001}
{Vanden Berk} D.~E.,  et~al., 2001, \mn@doi [AJ] {10.1086/321167}, 122, 549

\bibitem[\protect\citeauthoryear{Veale, White  \& Conroy}{Veale
  et~al.}{2014}]{Veale2014}
Veale M.,  White M.,   Conroy C.,  2014, \mn@doi [MNRAS]
  {10.1093/mnras/stu1821}, 445, 1144

\bibitem[\protect\citeauthoryear{Venemans, Walter, Zschaechner, Decarli, {De
  Rosa}, Findlay, McMahon  \& Sutherland}{Venemans
  et~al.}{2015}]{Venemans2015c}
Venemans B.~P.,  Walter F.,  Zschaechner L.,  Decarli R.,  {De Rosa} G.,
  Findlay J.~R.,  McMahon R.~G.,   Sutherland W.~J.,  2015, \mn@doi [ApJ]
  {10.3847/0004-637x/816/1/37}, 816, 37

\bibitem[\protect\citeauthoryear{Venemans et~al.,}{Venemans
  et~al.}{2017}]{Venemans2017}
Venemans B.~P.,  et~al., 2017, \mn@doi [ApJ] {10.3847/1538-4357/aa62ac}, 837,
  146

\bibitem[\protect\citeauthoryear{Vito et~al.,}{Vito et~al.}{2018}]{Vito2018}
Vito F.,  et~al., 2018, \mn@doi [MNRAS] {10.1093/mnras/stx2486}, 473, 2378

\bibitem[\protect\citeauthoryear{Volonteri \& Reines}{Volonteri \&
  Reines}{2016}]{Volonteri2016}
Volonteri M.,  Reines A.~E.,  2016, \mn@doi [ApJ] {10.3847/2041-8205/820/1/l6},
  820, L6

\bibitem[\protect\citeauthoryear{Warren, Hewett  \& Osmer}{Warren
  et~al.}{1994}]{Warren1994}
Warren S.~J.,  Hewett P.~C.,   Osmer P.~S.,  1994, \mn@doi [ApJ]
  {10.1086/173660}, 421, 412

\bibitem[\protect\citeauthoryear{Willott et~al.,}{Willott
  et~al.}{2010a}]{Willott2010}
Willott C.~J.,  et~al., 2010a, \mn@doi [AJ] {10.1088/0004-6256/139/3/906}, 139,
  906

\bibitem[\protect\citeauthoryear{Willott et~al.,}{Willott
  et~al.}{2010b}]{Willott2010a}
Willott C.~J.,  et~al., 2010b, \mn@doi [AJ] {10.1088/0004-6256/140/2/546}, 140,
  546

\bibitem[\protect\citeauthoryear{Willott, Omont  \& Bergeron}{Willott
  et~al.}{2013}]{Willott2013quasars}
Willott C.~J.,  Omont A.,   Bergeron J.,  2013, \mn@doi [ApJ]
  {10.1088/0004-637X/770/1/13}, 770, 13

\bibitem[\protect\citeauthoryear{Wyithe \& Loeb}{Wyithe \&
  Loeb}{2003}]{Wyithe2003}
Wyithe J. S.~B.,  Loeb A.,  2003, ApJ, 595, 614

\bibitem[\protect\citeauthoryear{Yang et~al.,}{Yang et~al.}{2020}]{Yang2020}
Yang J.,  et~al., 2020, \mn@doi [ApJ] {10.3847/2041-8213/ab9c26}, 897, L14

\bibitem[\protect\citeauthoryear{van~der Burg, Hildebrandt  \& Erben}{van~der
  Burg et~al.}{2010}]{VanderBurg2010}
van~der Burg R. F.~J.,  Hildebrandt H.,   Erben T.,  2010, \mn@doi [A{\&}A]
  {10.1051/0004-6361/200913812}, 523, A74

\makeatother
\end{thebibliography}


\appendix

\section{Spectroscopically confirmed sources}\label{sect:appendixA}

In Table~\ref{table:spec} we present the brightest sources in our parent sample that have been spectroscopically confirmed.
In addition to the published spectroscopic redshift, for some of these sources we were able to obtain reduced spectra which we present in Fig.~\ref{fig:spectra}.
As discussed further in~\citet{Adams2020} we overlap with the majority of the~\citet{Boutsia2018} sample.
As part of our analysis we identified a discrepancy between the absolute magnitudes presented by~\citet{Boutsia2018}, which can be up to $1\,{\rm mag}$ fainter than the $M_{\rm UV}$ we calculated from our best-fitting SED model.
In~\citet{Boutsia2018} the $M_{\rm UV}$ is estimated by applying a $K$-correction to the $r-$band data, which typically hosts the Lyman-break from $z = 3.9$--$4.7$.
Our analysis demonstrates that this method can underestimate the $M_{\rm UV}$. 
In addition to this discrepancy, we find we cannot reproduce the higher number density of AGN derived by~\citet{Boutsia2018} ($\phi = 1.6 \times 10^{-6}$ at $M_{\rm UV} = -23.5$).
This is puzzling given that the majority of the~\citet{Boutsia2018} sources are reselected in this work.

\begin{table*}
\caption{The spectroscopically confirmed high-redshift sources from the full $z \simeq 4$ sample.  We present the objects with $M_{\rm UV} < -22.0$, with the COSMOS and XMM-LSS sources in the upper and lower part of the table respectively.
The first column is the source ID number followed by the R.A. and Declination.
In column 3 we present the total HSC $I$-band apparent magnitude followed by the best-fitting photometric redshift with a Galaxy and QSO template in columns 4 and 5.
In brackets after the photometric redshift is the $\chi^2$ of the fit.
The final two columns denote the absolute UV magnitude followed by a note indicating the origin of the spectroscopic measurement.
KB18 corresponds to~\citet{Boutsia2018}, zCOS to zCOSMOS, PR to Primus, VAN to VANDELS.
We denote with an asterick the objects for which we were able to obtain the rest-frame UV spectrum, presented in Section~\ref{sect:spectra}.
}\label{table:spec}
\begin{tabular}{lcccccccl}
\hline
ID & R.A. & Dec. & $I_{\rm HSC}$ & $z_{\rm gal}$ & $z_{\rm qso}$ & $z_{\rm spec}$ & $M_{\rm UV}$ & Notes \\
\hline
188657 & 9:57:52.16 & +1:51:20.08 & 21.20 & 0.27 (39.0) & 3.92 (20.1) & 4.174 & -24.90 & KB18(658294),zCOS* \\
203718 & 10:01:56.55 & +1:52:18.80 & 21.88 & 4.19 (70.3) & 4.28 (50.3) & 4.447 & -24.18 & zCOS,PR* \\
657658 & 10:02:48.91 & +2:22:11.88 & 21.69 & 3.75 (96.0) & 3.52 (151.7) & 3.748 & -24.12 & KB18(1163086),zCOS,PR* \\
702265 & 10:00:24.23 & +2:25:09.86 & 22.55 & 4.28 (123.4) & 4.36 (98.0) & 4.596 & -23.94 & PR \\
153468 & 9:58:08.09 & +1:48:33.10 & 22.40 & 3.92 (19.1) & 3.88 (25.4) & 3.986 & -23.75 & KB18(664641) \\
113309 & 10:00:25.77 & +1:45:33.11 & 22.40 & 3.79 (126.7) & 4.16 (87.0) & 4.140 & -23.72 & KB18(330806) \\
677759 & 10:02:33.23 & +2:23:28.74 & 22.45 & 3.50 (123.0) & 3.72 (58.1) & 3.650 & -23.43 & KB18(1159815) \\
724788 & 9:59:06.46 & +2:26:39.39 & 22.35 & 3.73 (122.8) & 3.92 (81.7) & 4.170 & -23.42 & KB18(1273346),PR \\
523406 & 9:59:31.01 & +2:13:32.88 & 22.46 & 3.48 (70.0) & 3.60 (52.7) & 3.650 & -23.37 & KB18(1054048) \\
908052 & 9:59:22.37 & +2:39:32.63 & 23.49 & 3.72 (76.3) & 4.08 (32.9) & 3.748 & -22.99 & KB18(1730531) \\
840823 & 10:00:54.52 & +2:34:34.90 & 23.75 & 4.43 (15.1) & 4.56 (36.7) & 4.539 & -22.57 & DEIMOS(842313) \\
112866 & 10:01:26.67 & +1:45:26.16 & 23.87 & 4.44 (9.2) & 4.60 (11.6) & 5.137 & -22.49 & DEIMOS(308643) \\
654636 & 10:01:31.60 & +2:21:57.73 & 24.01 & 4.48 (8.6) & 4.52 (25.1) & 4.511 & -22.34 & VUDS(5101210235) \\
606304 & 10:01:12.50 & +2:18:52.58 & 24.09 & 4.46 (60.9) & 4.44 (32.0) & 5.691 & -22.27 & VUDS(5101218326) \\
697618 & 10:01:19.91 & +2:24:47.47 & 24.21 & 4.46 (2.3) & 4.56 (8.4) & 4.419 & -22.14 & DEIMOS(733857) \\
\hline
278034 & 2:18:44.46 & -4:48:24.59 & 19.74 & 4.13 (144.0) & 4.44 (98.3) & 4.574 & -26.48 & SDSS* \\
1364622 & 2:27:54.62 & -4:45:35.37 & 20.31 & 3.60 (48.6) & 3.72 (97.3) & 3.741 & -25.64 & SDSS* \\
919928 & 2:24:13.41 & -5:27:24.73 & 20.31 & 3.56 (44.2) & 3.80 (24.2) & 3.779 & -25.55 & SDSS* \\
1448906 & 2:25:27.23 & -4:26:31.21 & 21.32 & 3.54 (66.1) & 3.68 (29.3) & 3.835 & -24.57 & PR, VVDS* \\
45737 & 2:18:05.65 & -5:26:35.58 & 21.75 & 3.77 (45.5) & 3.64 (102.1) & 4.077 & -24.21 & PR \\
456332 & 2:17:14.17 & -4:20:00.54 & 22.30 & 0.47 (65.4) & 4.36 (52.2) & 4.317 & -24.12 & PR \\
307263 & 2:18:31.37 & -4:43:54.39 & 21.88 & 3.42 (32.0) & 3.64 (13.9) & 3.683 & -24.03 & PR \\
1448401 & 2:27:54.45 & -4:26:37.97 & 22.29 & 3.63 (24.3) & 3.64 (66.4) & 3.835 & -23.62 & VVDS* \\
173860 & 2:17:34.38 & -5:05:14.55 & 23.25 & 0.35 (63.9) & 3.88 (28.2) & 3.983 & -22.72 & VAN(199159)* \\
75407 & 2:17:53.11 & -5:21:24.40 & 23.67 & 0.35 (14.0) & 4.20 (10.0) & 3.802 & -22.64 & VAN(141491)* \\
1499379 & 2:25:33.71 & -4:15:41.51 & 23.68 & 0.42 (21.7) & 4.28 (16.5) & 3.699 & -22.62 & VVDS* \\
90440 & 2:18:05.17 & -5:18:55.74 & 23.67 & 0.41 (15.5) & 4.16 (9.6) & 3.921 & -22.58 & VAN(150302)* \\
1463494 & 2:27:53.87 & -4:23:20.34 & 23.37 & 3.36 (69.4) & 3.56 (54.2) & 3.626 & -22.42 & VVDS* \\
1485639 & 2:26:59.61 & -4:18:32.88 & 23.67 & 3.82 (10.1) & 4.08 (15.8) & 3.872 & -22.36 & VVDS* \\
140721 & 2:17:33.77 & -5:10:24.57 & 23.76 & 3.98 (13.4) & 4.20 (22.3) & 4.129 & -22.33 & VAN(018574)* \\
1522259 & 2:25:33.61 & -4:10:57.99 & 23.79 & 4.12 (20.6) & 4.28 (15.9) & 4.116 & -22.32 & VVDS* \\
\hline
\end{tabular}
\end{table*}%


\section{AGN fraction from our fitting}\label{sect:appendixB}

\begin{figure}
\begin{center}
\includegraphics[width =0.48\textwidth]{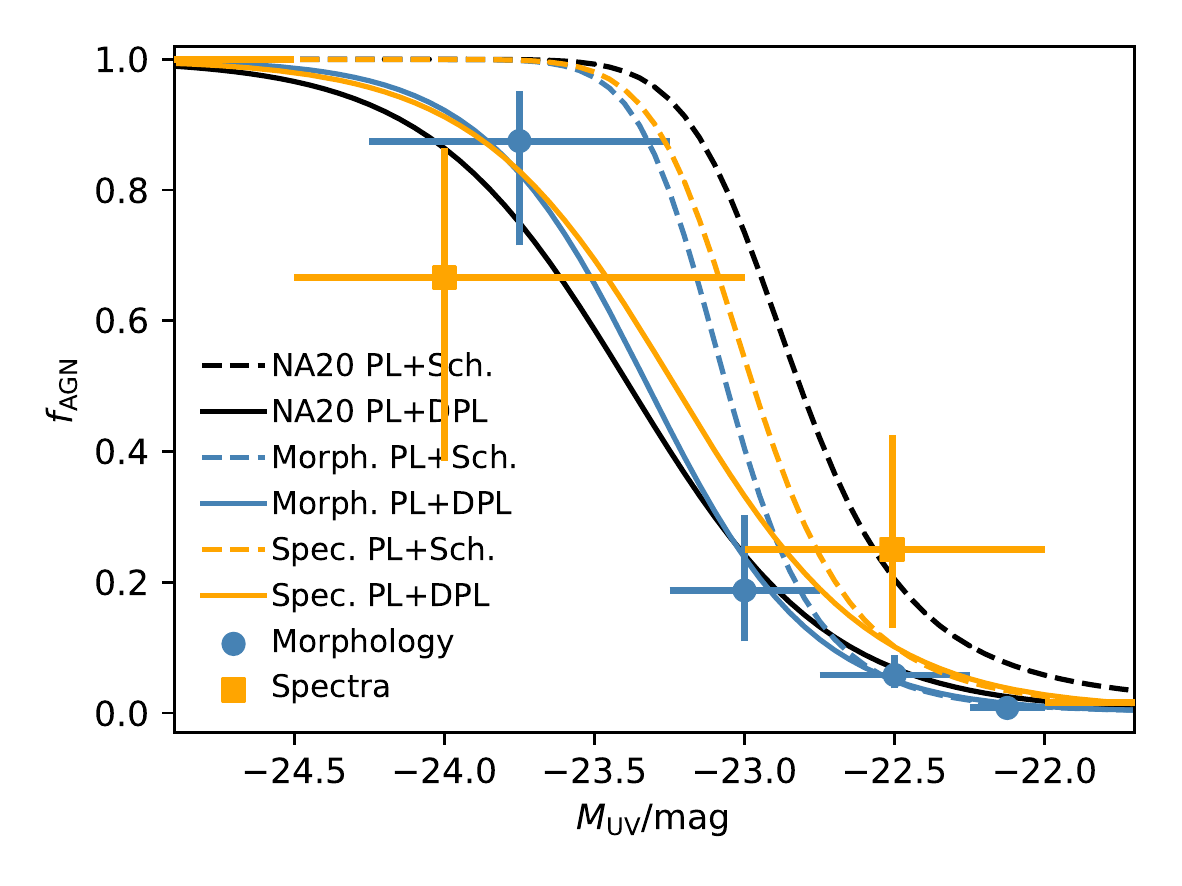}
\caption{The results of our fitting to the separated AGN and SF-dominated sources.  The solid (dashed) lines show the $f_{\rm AGN}$ derived when assuming a DPL (Schechter) function for the SF component.  In both cases a power-law is assumed for the AGN component.
The black lines show the results from a simultaneous fit of both types of sources by~\citet{Adams2020}.
The blue (orange) lines show the results from this work, where we have separated SF and AGN-dominated sources according to the morphology (spectra).
We reproduce our observed $f_{\rm AGN}$ points as in Fig.~\ref{fig:agnfrac}.
}
\label{fig:agnfracsimple}
\end{center}
\end{figure}

In Fig.~\ref{fig:agnfracsimple} we show the resulting AGN fractions derived from our fitting procedure to the separated AGN and SF UV LFs.
We compare these to the results from~\citet{Adams2020} who did not separate the two populations.
In comparison to~\citet{Adams2020} we find that the DPL and Schechter parameterisations of the SF-dominated sources are consistent within $1\sigma$.
The observed differences in the $f_{\rm AGN}$ are instead found to be due to more significant changes in the derived parameters for the power-law fit to the AGN-dominated sub-sample.
When a Schechter function form is assumed for the SF-dominated sources, the fit to the AGN-dominated objects has a shallower slope than in~\citet{Adams2020}, leading to a brighter point of transition.
Conversely, when a DPL function form is assumed for the SF-dominated sources, the power-law fit to the AGN-dominated objects shows a higher normalisation and hence the transition point (where $f_{\rm AGN} \simeq 0.5$) moves faint-wards.

\bsp	
\label{lastpage}
\end{document}